\documentclass[%
reprint,
superscriptaddress,
amsmath,amssymb,
aps,
pra,
]{revtex4-2}

\usepackage{graphicx} % include figure files
\usepackage{dcolumn} % align table columns on decimal point
\usepackage{bm} % bold math
\usepackage{bbm}
\usepackage{braket}
\usepackage{comment}
\usepackage[dvipsnames]{xcolor}

\usepackage[resetlabels,labeled]{multibib}
\usepackage{lipsum}

\usepackage{booktabs}

\newcites{First}{References}

\definecolor{darkblue}{rgb}{0.,0.,0.6}
\definecolor{darkgreen}{rgb}{0.,0.6,0.0}
\definecolor{darkorange}{rgb}{0.8,0.4,0.2}

\newcommand{\Yb}{$^{171}$Yb$^+$\,}
\newcommand{\Ybc}{$^{172}$Yb$^+$\,}
\usepackage{hyperref} % add hypertext capabilities

\begin{document}
\preprint{APS/123-QED}

% title
\title{Quantum Simulation of Charge and Exciton Transfer \\in Multi-mode Models using Engineered Reservoirs}

% author list
\author{Visal So}
\email{vs39@rice.edu}
\affiliation{Department of Physics and Astronomy and Smalley-Curl Institute, Rice University, Houston, TX 77005, USA}
\author{Midhuna Duraisamy Suganthi}
\affiliation{Department of Physics and Astronomy and Smalley-Curl Institute, Rice University, Houston, TX 77005, USA}
\affiliation{Applied Physics Graduate Program, Smalley-Curl Institute, Rice University, Houston, TX 77005, USA }
\author{Mingjian Zhu}
\affiliation{Department of Physics and Astronomy and Smalley-Curl Institute, Rice University, Houston, TX 77005, USA}
\author{Abhishek Menon}
\affiliation{Department of Physics and Astronomy and Smalley-Curl Institute, Rice University, Houston, TX 77005, USA}
\author{George Tomaras}
\affiliation{Department of Physics and Astronomy and Smalley-Curl Institute, Rice University, Houston, TX 77005, USA}
\affiliation{Applied Physics Graduate Program, Smalley-Curl Institute, Rice University, Houston, TX 77005, USA }
\author{Roman Zhuravel}
\affiliation{Department of Physics and Astronomy and Smalley-Curl Institute, Rice University, Houston, TX 77005, USA}
\author{Han Pu}
\affiliation{Department of Physics and Astronomy and Smalley-Curl Institute, Rice University, Houston, TX 77005, USA}
\author{Peter G. Wolynes}
\affiliation{Department of Physics and Astronomy and Smalley-Curl Institute, Rice University, Houston, TX 77005, USA}
\affiliation{Department of Chemistry, Rice University, Houston, TX 77005, USA}
\affiliation{Center for Theoretical Biological Physics, Rice University, Houston, TX 77005, USA}
\affiliation{Department of Biosciences, Rice University, Houston, TX 77005, USA}
\author{José N. Onuchic}
\affiliation{Department of Physics and Astronomy and Smalley-Curl Institute, Rice University, Houston, TX 77005, USA}
\affiliation{Department of Chemistry, Rice University, Houston, TX 77005, USA}
\affiliation{Center for Theoretical Biological Physics, Rice University, Houston, TX 77005, USA}
\affiliation{Department of Biosciences, Rice University, Houston, TX 77005, USA}
\author{Guido Pagano}
\email{pagano@rice.edu}
\affiliation{Department of Physics and Astronomy and Smalley-Curl Institute, Rice University, Houston, TX 77005, USA}

%----------------------------
% abstract
%----------------------------
\begin{abstract}

Quantum simulation offers a route to study open‐system molecular dynamics in non-perturbative regimes by programming the interactions among electronic, vibrational, and environmental degrees of freedom on similar energy scales. Trapped-ion systems possess this capability, with their native spins, phonons, and tunable dissipation integrated within a single platform. Here, we demonstrate an open-system quantum simulation of charge and exciton transfer in a \emph{multi-mode} linear vibronic coupling model. Employing tailored spin-phonon interactions alongside reservoir engineering techniques, we emulate a system with two dissipative vibrational modes coupled to donor and acceptor electronic sites and follow its non-equilibrium dynamics. We continuously tune the system from the charge transfer (CT) regime to the vibrationally assisted exciton transfer (VAET) regime by controlling the vibronic coupling strengths. We find that degenerate modes enhance CT and VAET rates at large energy gaps, while non-degenerate modes activate slow-mode pathways that reduce the energy-gap dependence, thus enlarging the window for efficient transfer. These results show that the presence of one additional vibration introduces interfering vibrationally assisted pathways and reshapes non-perturbative quantum excitation transfer. Our work establishes a scalable and hardware-efficient route to simulating chemically relevant, many-mode vibronic processes with engineered environments, guiding the design of next-generation organic photovoltaics and molecular electronics.

\end{abstract}

%----------------------------
% display title
%----------------------------
\maketitle

%----------------------------
% introduction
%----------------------------

Molecular vibrations drive a wide range of phenomena in charge and energy transfer in complex chemical and biological systems. Understanding these processes requires modeling the interactions among the electronic, spin, and vibrational degrees of freedom, which cannot be treated independently---especially when the Born-Oppenheimer (BO) approximation breaks down. Although the BO approximation is the cornerstone of structural chemistry, it fails in cases where nuclear and electronic motions become strongly coupled, as occurs in nitrogen fixation and photosynthesis. The fully simultaneous quantum treatment of electrons and vibrations is still a daunting task for existing numerical methods \cite{behrman1983montecarlo, tanimura1992interplay, kundu2022eet, scholes2025challenge, zhang2024quantum}.

Quantum rate phenomena generally require a model that accounts for both fast, intramolecular vibrational modes and slower, environment-mediated modes. Linear vibronic coupling models (LVCMs) offer the simplest framework for describing such systems by assuming that electronic states couple to multiple vibrational modes only in a linear fashion \cite{fassioli2014vibration,bakulin2015organic,kang2024chemical}. LVCMs have been widely used to model many complex processes, including singlet fission, triplet-triplet annihilation, and charge transfer in organic photovoltaics \cite{johnson1992vibration,rather2020vibrational,arsenault2021vibronic}.

The high degree of control and tunability of programmable quantum platforms---such as trapped ions \cite{kang2024chemical}, superconducting qubits \cite{Dutta2024}, and photonic devices \cite{sparrow2018}---potentially provides an alternative approach to large-scale classical computations for studying condensed-phase chemical dynamics through direct quantum simulation. 
In particular, trapped-ion analog and analog-digital simulators have already realized a variety of chemical dynamics with fully programmable system parameters and accurate time-resolved features \cite{gorman2018VAET,valahu2023conical,whitlow2023conical,macdonell2023timedomain,ke2023plet}.
With engineered reservoirs on their native spin and bosonic degrees of freedom, trapped-ion systems have also been recently used to study non-equilibrium quantum reaction dynamics  \cite{maier2019enaqt, so2024electrontransfer,sun2024quantumsimulationspinbosonmodels, navickas2024experimentalquantumsimulationchemical,pagano2025varenna}.

\begin{figure}
    \centering
    \includegraphics[width=0.9\linewidth]{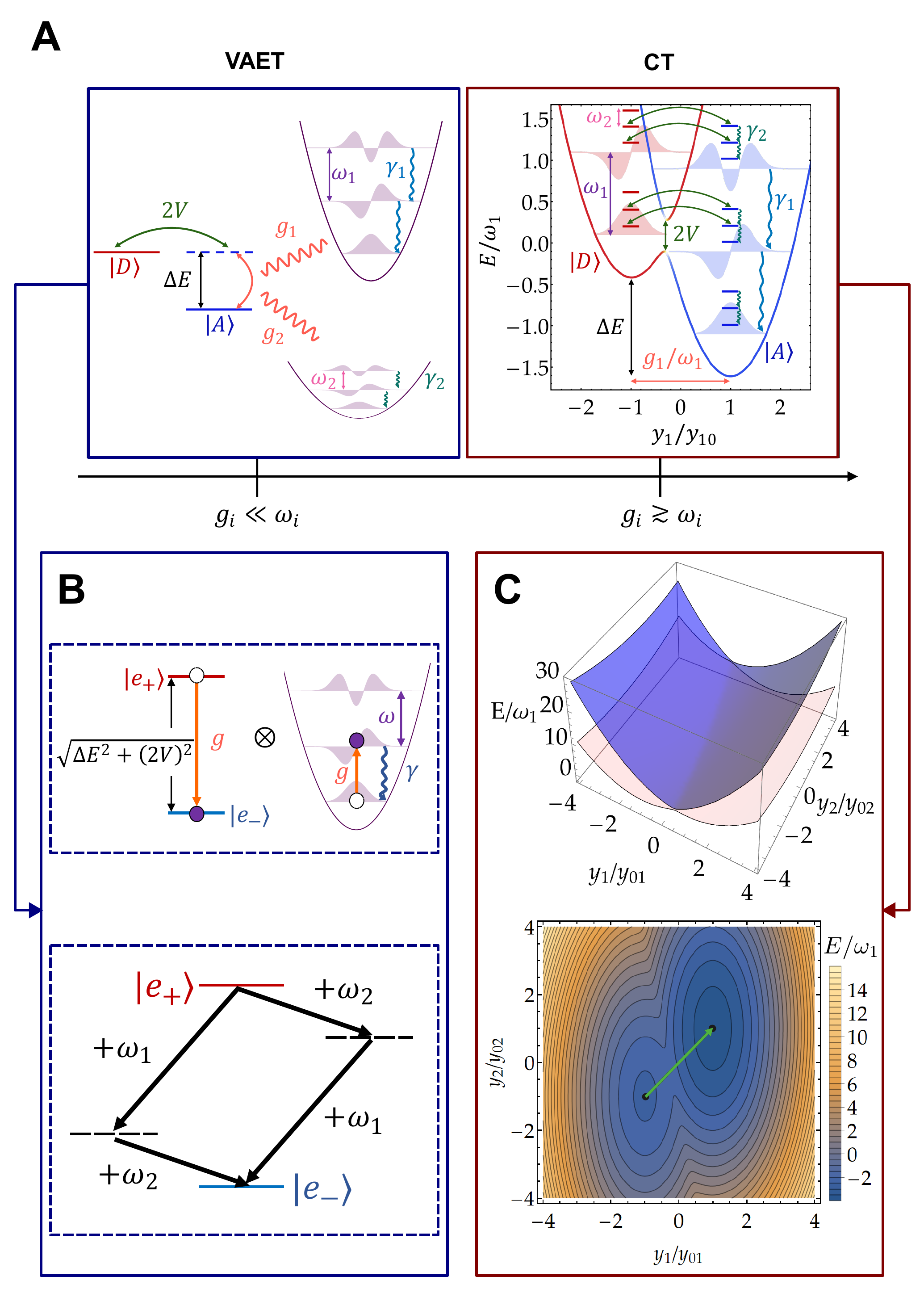}
    \caption{{\bf Two-mode LVCM with engineered reservoirs.} 
    ({\bf A}) Schematic diagram illustrating two regimes of transfer dynamics, defined by vibronic coupling strengths. In the VAET regime, the harmonic oscillators are weakly coupled to the donor–acceptor electronic system, whereas the strong vibronic couplings in the CT regime introduce significant displacements, resulting in distinct two-dimensional potential energy surfaces for the donor (red) and acceptor (blue) states. The electronic coupling opens a $2V$ avoided crossing between these surfaces. For clarity, we show only the $y_2 = 0$ cut across the two-dimensional potential energy surfaces, with the vibrational levels associated with the fast $(\omega_1)$ and slow ($\omega_2$) modes, represented by their wavefunctions and horizontal lines, respectively. In both regimes, the vibrational modes undergo dissipation, represented by the wiggly arrows. \textbf{(B)} Top (dashed panel): Illustration of a first-order VAET process at zero temperature, where weak vibronic coupling coherently drives transitions between the eigenstates of the electronically coupled system via a single-excitation exchange with a vibrational mode (see Appendix \ref{SM_weakcouplingVAET}). Specifically, when energy is released from the higher-energy eigenstate, the vibrational mode gains a phonon excitation, which subsequently dissipates into the environment. Bottom (dashed panel): Two constructively interfering transfer pathways underlying the second-order VAET process, which bridges the energy gap of $\sqrt{\Delta E^2 + (2V)^2} = \omega_1 + \omega_2$. \textbf{(C)} Top: Two-dimensional energy landscape in the CT regime, with coordinates defined by the two vibrational modes. The blue and pink surfaces correspond to the system's upper and lower adiabatic surfaces under strong electronic coupling, respectively. Bottom: Contour plot of the lower adiabatic surface, with the green arrow indicating the direction of charge transfer along the effective reaction coordinate.}
    \label{fig_concept}
\end{figure}

In this work, we present the first trapped-ion simulation in which unitary spin–phonon couplings and mode-selective dissipation via cooling are independently programmed, enabling real-time open-system emulation of excitation transfer dynamics in a two-mode LVCM. Engineering spin-phonon systems with multiple bosonic modes, combined with environment engineering, represents an important step toward realizing models of molecular systems in which multiple intramolecular (fast) modes coexist with long-wavelength, low-frequency environment-mediated modes---both of which play crucial roles in excitation-energy transfer \cite{Renger2012}. Here, by using both ground-state and optical qubits, we achieve simultaneous control over mode frequencies, vibronic couplings, and dissipation rates. We characterize the low-temperature transfer rate across a wide parameter range, highlighting the roles of mode degeneracy and vibronic coupling strength.
We focus on two transfer regimes (see Fig.~\ref{fig_concept}A): charge transfer (CT) and vibrationally assisted exciton transfer (VAET), which correspond to strong and weak vibronic coupling, respectively. The distinction is made based on the differing phenomenological effects of the vibronic coupling on typical excitation and charge transfer processes \cite{fassioli2014vibration,chen2015optimal}. When the vibrational modes are strongly coupled to the electronic states of the system, they reshape the potential energy landscape and actively influence the charge transfer reaction \cite{schlawin2021electrontransfer}. By contrast, weak couplings between the vibrational and electronic degrees of freedom primarily allow phonons to assist coherence between the electronic eigenstates \cite{li2021multimodeVAET}. We note that charge transfer can also occur in the weak vibronic coupling regime \cite{matyushov2023reorganizationenergy}. However, in this study, we use this term to refer specifically to the strong vibronic coupling regime, with which it is more commonly associated.

%----------------------------
% model
%----------------------------

The minimal multi-mode LVCM considers two vibrational modes and is described by the following Hamiltonian ($\hbar = 1$) \cite{onuchic1987twomode}:
\begin{equation}
    H \!= V\sigma_x\! + \frac{\Delta E}{2}\sigma_z \!+ \sum_{i=1}^2\left\{\frac{g_i}{2}\sigma_z\!\left(a_i+a^\dag_i\right)\!+\omega_i a^\dag_i a_i\right\},
    \label{eq_H2M}
\end{equation}
where $\sigma_{x,z}$ are the Pauli operators acting on the donor and acceptor electronic sites, $\ket{D}\equiv\ket{\uparrow}_z$ and $\ket{A}\equiv\ket{\downarrow}_z$, with an energy difference $\Delta E$ and an electronic coupling strength $V$. For $\Delta E > 0$, the Hamiltonian describes an exothermic reaction when the excitation is transferred from the donor site to the acceptor site. Each harmonic oscillator $i$ with a vibrational energy $\omega_i$ is associated with creation ($a_i^\dagger$) and annihilation ($a_i$) operators, and it is linearly coupled to the electronic sites at a rate $g_i$.

In this model, the charge transfer (CT) regime is realized when the vibronic coupling is comparable to or larger than the harmonic frequency ($g_i\gtrsim\omega_i$, also known as strong coupling in quantum optics), which is characteristic of many situations involving charge motion in chemical and biological reactions, ranging from redox catalysis to solvent-induced electronic delocalization \cite{paul1996et,barthel2001solvent,hsu2020chargetransfer,schlawin2021electrontransfer}. The donor and acceptor electronic sites in this regime are described by uncoupled two-dimensional potential energy surfaces with respect to the ($y_1$, $y_2$) spatial coordinates, where $y_i = y_{i0}(a_i+a_i^\dagger)/2$ with $y_{i0}=\sqrt{1/2m\omega_i}$ and $m$ being the particle mass. The electronic coupling $V$ mixes the two potential energy surfaces by opening an avoided crossing with a gap of $2V$ at their intersection. A strong vibronic coupling ($g_i\gtrsim\omega_i$) distorts the energy landscape by inducing a displacement of $g_i/\omega_i$ between the donor and acceptor potential energy surfaces along $y_i$, providing the electronic coupling with a dependence on the overlaps of the displaced oscillator states \cite{skourtis1992photosynthesis,schlawin2021electrontransfer,so2024electrontransfer}.
During CT, the donor population undergoes crossing of the energy barrier along an effective reaction coordinate, roughly defined by the axis $y_1 + y_2$ in the degenerate ($\omega_1=\omega_2$) case. The total reorganization energy of the system is defined as $\lambda= \sum_i\lambda_i\equiv\sum_ig_i^2/\omega_i$, which is the energy required to displace the wave packet on an uncoupled potential energy surface by the distances of $|g_1/\omega_1|$ and $|g_2/\omega_2|$ along the $y_1$ and $y_2$ axes, respectively. When the electronic coupling is small ($V\approx0$), the reorganization energy $\lambda$ and energy difference between the two surfaces $\Delta E$ determine the classical activation energy, defined by $U = (\Delta E +\lambda)^2/4\lambda$ \cite{schlawin2021electrontransfer,so2024electrontransfer}.

Within the CT regime, there exist two electronic coupling regimes: the strictly nonadiabatic (or perturbative) regime with $|V|<\lambda_i/4$ and the strongly adiabatic (or non-perturbative) regime with $|V|\sim\lambda_i/4$ \cite{schlawin2021electrontransfer,so2024electrontransfer,padilla2025delocalizedexcitationtransferopen}. In the former, Fermi's golden rule---treating the electronic coupling as a perturbation to the uncoupled donor-acceptor system---predicts that transfer dynamics occur when the quantized vibronic energy levels of the donor and acceptor surfaces match, resulting in transfer rate resonances at:

\begin{equation}
    \Delta E \approx \ell_1\omega_1 + \ell_2\omega_2,
    \label{eq_nonadiabatic}
\end{equation}
with $\ell_1$ and $\ell_2$ being integers (see Appendix \ref{SM_weakcouplingET}). On the other hand, at larger electronic coupling ($|V|\sim\lambda_i/4$), the avoided crossing becomes appreciable, leading to the hybridization of the two-dimensional donor and acceptor potential energy surfaces into upper and lower adiabatic energy surfaces (see Fig.~\ref{fig_concept}C). This results in delocalized eigenstates that are superpositions of donor and acceptor vibronic states  \cite{schlawin2021electrontransfer,so2024electrontransfer}. In this regime, Fermi's golden rule no longer applies, which provides strong motivation to study these transfer conditions experimentally.

\begin{figure}
    \centering
    \includegraphics[width=0.85\linewidth]{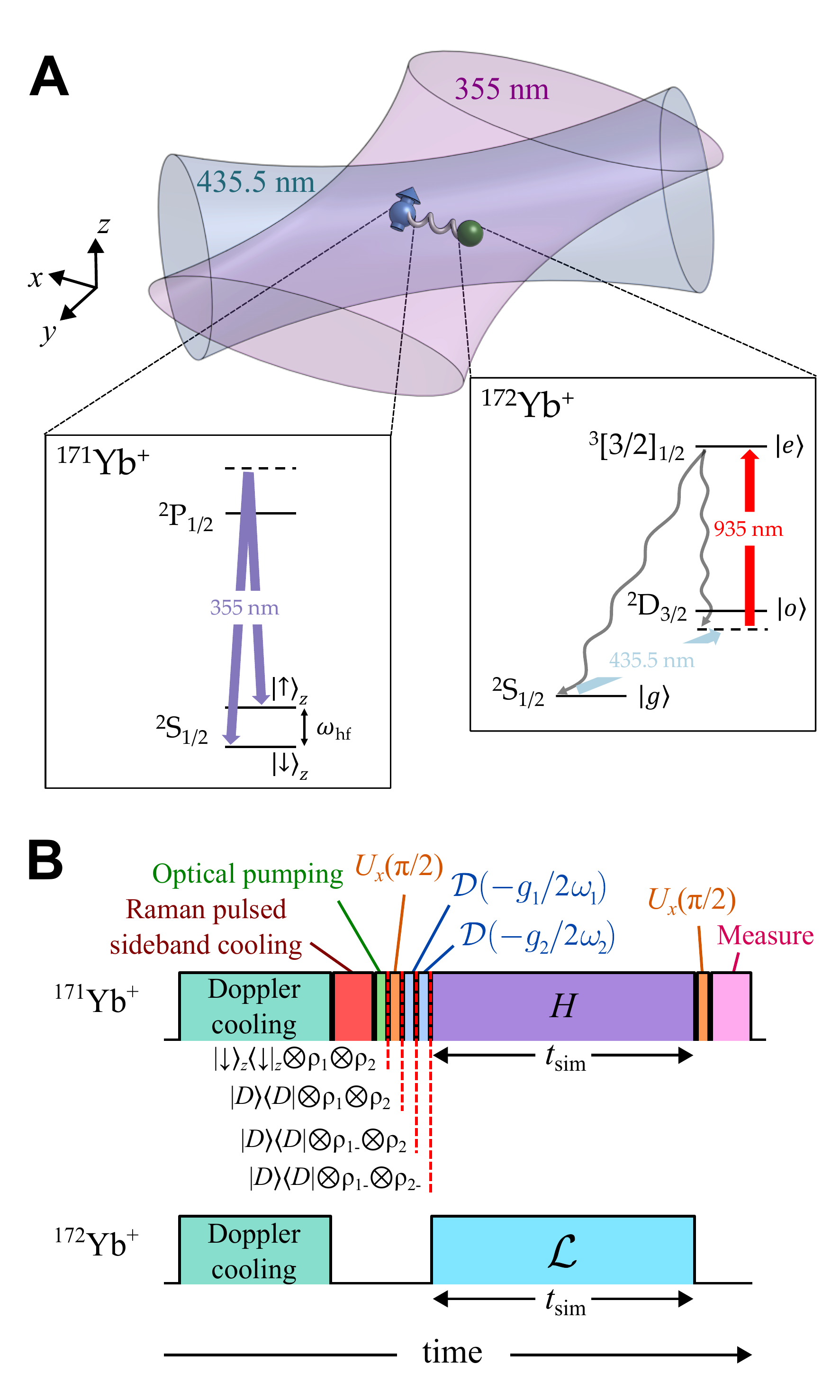}
    \caption{{\bf Trapped-ion quantum simulation of a two-mode LVCM with engineered reservoirs.} 
    ({\bf A}) Experimental setup for studying open-system LVCM dynamics with two vibrational modes using a \Yb-\Ybc ion chain, whose motional degrees of freedom are shared (represented by the connecting spring). Insets: simplified level schemes for \Yb and \Ybc qubits. Stimulated Raman transitions on the \Yb ground-state qubit with 355 nm beams (purple) are used to engineer the Hamiltonian in Eq.~\eqref{eq_H2M}. The optical qubit of \Ybc is addressed with a 435.5 nm beam (light blue) and a 935 nm repumper beam (red line in the \Ybc inset) for sympathetic cooling. ({\bf B}) Experimental sequence used to measure the time-resolved evolution of the excitation at the donor electronic site (see Methods for details). Each dashed red line indicates the state of the system after each preparation pulse.}
    \label{fig_experiment}
\end{figure}

Conversely, weak vibronic coupling ($g_i\ll\omega_i$) is more characteristic of vibrationally assisted exciton transfer (VAET), as occurs between pigments in light-harvesting compounds and their reaction centers \cite{gorman2018VAET}. In this regime, excitation transfer between donor and acceptor sites with an energy separation $\Delta E$ is enabled by the electronic coupling term $V\sigma_x$, resulting in two eigenenergies of the total electronic system that differ by $\sqrt{\Delta E^2 +(2V)^2}$.
Unlike in the CT regime, where strong vibronic couplings displace the donor and acceptor energy surfaces and thereby define the vibronic states that participate in excitation transfer, the couplings between the electronic system and the vibrational modes in the VAET regime are weak. From Fermi's golden rule, these weak couplings perturbatively facilitate quantized energy exchange between the oscillators and the electronically coupled excitation sites, leading to transfer resonances at:
\begin{equation}
\Delta E\approx\sqrt{(\ell_1\omega_1+\ell_2\omega_2)^2-(2V)^2},
\label{eq_VAETcondition}
\end{equation}
with $\ell_1$ and $\ell_2$ being integers. 
At resonance, the combined vibrational energy $\ell_1\omega_1+\ell_2\omega_2$ provided by the two modes exactly bridges the electronic energy gap, thereby assisting the transfer \cite{gorman2018VAET,sun2024quantumsimulationspinbosonmodels,li2021multimodeVAET} (see Appendix \ref{SM_weakcouplingVAET}).

%----------------------------
\begin{figure*}[t!]
    \centering\includegraphics[width=1\linewidth]{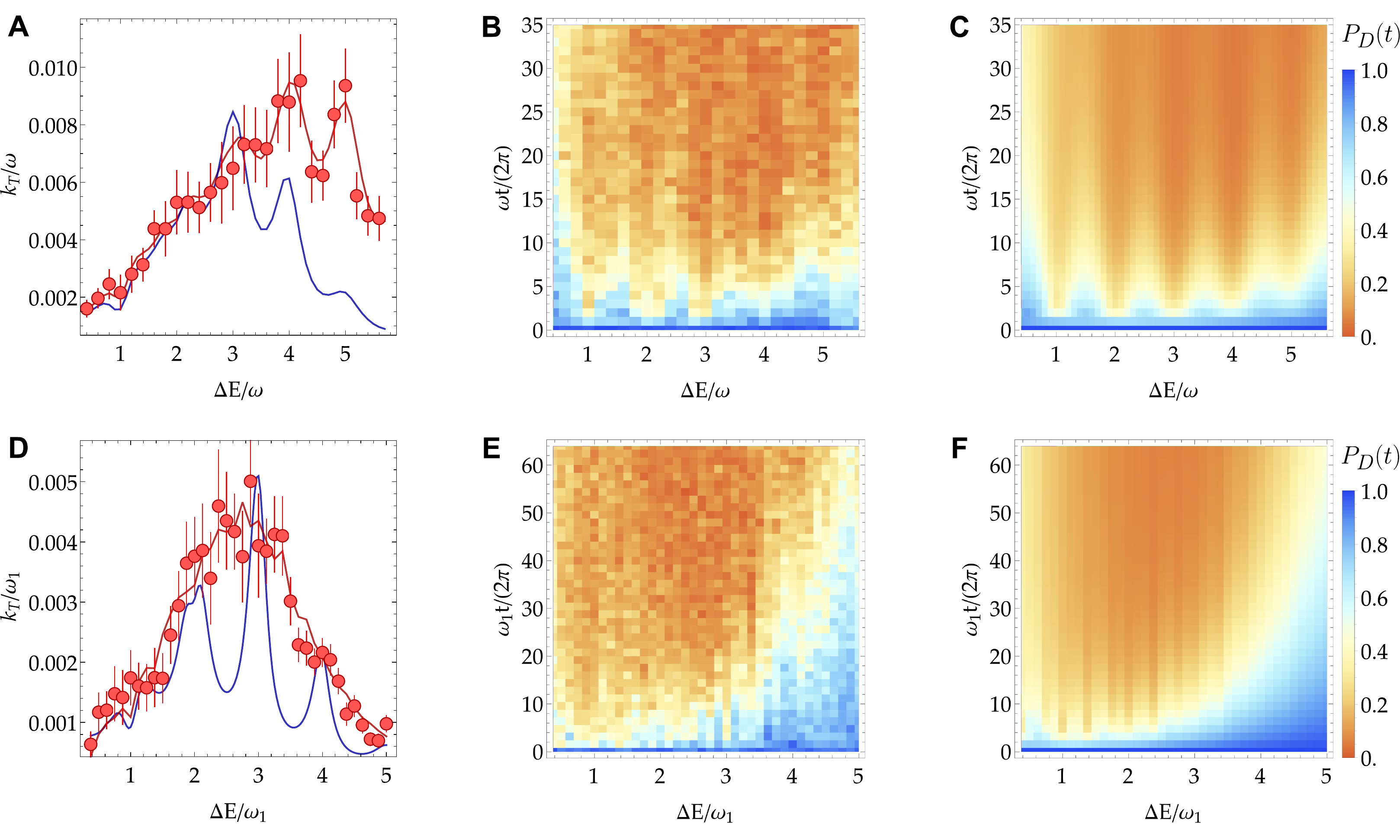}
    \caption{{\bf Transfer rates in the CT regime.} ({\bf A}) Transfer rate spectrum of degenerate CT ($\omega_1\!=\!\omega_2\!\equiv\!\omega$) with $(V,g_{1},\gamma_{1},g_{2},\gamma_{2})=(0.200, 1.200, 0.036, 1.100,0.040)\omega$. Red circles represent experimental data with error bars estimated via bootstrapping (see Methods). The solid red curve shows the transfer rate calculated from Eq.~\eqref{eq_master}, using the definition in Eq.~\eqref{eq_lifetime_kT} and including spin decoherence ($\gamma_{z}\!=\!0.0014 \omega$) and motional dephasing of both modes ($\gamma_{im}\!=\!0.0160\omega$, with $i=1,2$). The blue curve shows the numerical result for the single-mode CT case, where $\omega_2\!=\!g_2\!=\!\gamma_2\!=\!\gamma_{2m}\!=\!0$.
    ({\bf B and C}) Experimental and numerical donor population evolution $P_D(t)$ versus energy gap $\Delta E$ and the number of vibrational oscillations $\omega t/2\pi$, using the same parameters as in the red circles and solid red curve in {(A)}, respectively. Here, the detunings from the two tilt modes for encoding the degenerate vibrational energies are both set to $\delta_1=\delta_2=-2\pi\times5$ kHz. ({\bf D to F}) Same layout as in {(A to C)} for the non-degenerate CT case ($\omega_1\!>\!\omega_2$) with $(V,g_{1},\gamma_{1},\omega_{2},\gamma_{z},\gamma_{im})=(0.138, 1.029, 0.023, 0.375, 0.0009, 0.0100)\omega_{1}$ and $(g_{2},\gamma_{2})\!=\!(0.993,0.027)\omega_2$. The characteristic timescale is defined by the fast vibrational frequency $\omega_1$. In this case, the detuning from the inner (outer) tilt mode, which encodes the fast (slow) vibrational energy, is set to $\delta_1\!=\!-2\pi\!\times\!8$ kHz ($\delta_2\!=\!-2\pi\!\times\!3$ kHz).
    }
    \label{fig_CT}
\end{figure*}

In most realistic chemical situations, the interaction between the donor-acceptor vibronic system and the external environment causes the vibrational modes to undergo incoherent dissipation. Thus, the full Hamiltonian, $H_{\rm total}=H + H_{\rm b} + H_{\rm sb}$, also includes the bath degrees of freedom $H_{\rm b}$, described by a large collection of continuous harmonic oscillators, as well as a linear coupling between the bath and the system's vibrational modes, given by $H_{\rm sb}$. The correlations of the environment and their influence on the system can be characterized by a continuous spectral density function $J(\omega)$. Under the assumptions of a Markovian environment with Ohmic spectral densities \cite{garg1985friction, schlawin2021electrontransfer} and vibrational dissipation rates that are weaker than the vibrational and thermal energies ($\gamma_i \ll \omega_i, \gamma_i \ll k_B T_i$) \cite{lemmer2018engineering}, the dissipative dynamics of the system can be effectively described by a Lindblad master equation:
\begin{equation}
    \frac{\partial\rho}{\partial t}\!=-i[H,\rho] +\sum_{i=1}^2\left\{\gamma_i (\bar{n}_i\!+\!1)\mathcal{L}_{a_i}[\rho] + \gamma_i \bar{n}_i \mathcal{L}_{a^\dagger_i}[\rho]\right\},
    \label{eq_master}   
\end{equation}
where $\mathcal{L}_{c}[\rho]=
    c\rho c^\dagger - \frac{1}{2}\{c^\dagger c,\rho\}$, and $\bar{n}_i$ is the average thermal phonon number that describes the temperature of the environment for each vibrational mode $\omega_i$ with $k_B T_i \approx \omega_i/\log(1+1/\bar{n}_i)$. The vibrational excitations continuously evolve to equilibrate with the bath at a rate $\gamma_i$, leading to an irreversible population transfer from the donor site to the acceptor site \cite{schlawin2021electrontransfer,so2024electrontransfer}.

%----------------------------
% results
%----------------------------

% brief system description
To experimentally realize the multi-mode LVCM, we use a dual-species chain of one \Yb ion and one \Ybc ion trapped in a linear Paul trap \cite{so2024electrontransfer}. We encode the electronic degree of freedom in the two hyperfine clock states of the \Yb ground-state qubit, $\ket{^2S_{1/2}, F=1,m_F=0}\equiv\ket{\uparrow}_z$ and $\ket{^2S_{1/2}, F=0,m_F=0}\equiv\ket{\downarrow}_z$ with the frequency splitting of $\omega_\text{hf} = 2\pi \times 12.642$ GHz (see Fig.~\ref{fig_experiment}A). 
In this work, we use the two radial out-of-phase modes (also referred to as tilt modes), each along an orthogonal radial principal trap axis, to represent $\omega_1$ and $\omega_2$ (see Methods).
%-------------------------

% hamiltonian engineering
Our experiment is performed in a driven rotating frame, where we use two $\pi/2$ pulses to map the $z$ spin basis in Eq.~\eqref{eq_H2M} onto the $y$ basis. This method allows us to use the ion-light interactions between the \Yb qubit and the 355 nm Raman laser tones to independently engineer individual terms of the Hamiltonian in Eq.~\eqref{eq_H2M} with precise control. Two laser tones resonant with the qubit frequency generate the single-qubit operations ($V\sigma_x$ and $({\Delta E}/{2})\,\sigma_z$ terms), while two other laser tone pairs at frequencies $\pm\mu_i=\pm(\omega_{\text{tilt},i}+\delta_i)$ from the qubit resonance realize the vibronic coupling and harmonic terms in Eq.~\eqref{eq_H2M}. Here, $\delta_i$ is the detuning from its respective radial tilt mode at frequency $\omega_{\text{tilt},i}$, which determines the vibrational energy $\omega_i$ through $\delta_i \equiv -\omega_i$ \cite{schneider2012spinboson, so2024electrontransfer}.

% dissipation engineering
The engineered dissipation of the vibrational modes is realized by driving the narrow transitions from the ground-state manifold $\ket{g}\!\equiv\!\ket{^2S_{1/2}}$ to the optical metastable-state manifold $\ket{o}\equiv\ket{^2D_{3/2}}$ of the \Ybc ion, detuned by $-\omega_{\text{tilt},i}$ from resonance, using a total of four laser tones to address all the involved Zeeman sub-levels. Together with a 935 nm repumper beam to optically reset the electronic excitation to the ground states, this results in sympathetic cooling on both radial tilt modes of the chain with independently tunable dissipation rates. This is equivalent to generating a structured bath of continuous harmonic oscillators with two Lorentzian spectral densities centered at $\omega_1$ and $\omega_2$, with full widths at half maximum $\gamma_1$ and $\gamma_2$, respectively \cite{lemmer2018engineering, sun2024quantumsimulationspinbosonmodels} (see Appendix \ref{SM_niba}).
In this setup, all the system parameters---including the bath properties---are determined by the frequency and power of the laser tones used to generate the corresponding ion-light interactions. Therefore, they can be precisely tuned and independently calibrated \cite{so2024electrontransfer}, with the magnitudes of the system parameters used in our study being much larger than the decoherence rates caused by experimental imperfections (see Appendix \ref{SM_numeric}). The experimental sequence is discussed in Methods and summarized in Fig.~\ref{fig_experiment}B.

In this work, we focus on the non-perturbative quantum regime \cite{schlawin2021electrontransfer}, where the electronic coupling strength is strong ($|V|\!\sim\!\lambda_i/4$), and the bath temperatures are low ($\bar{n}_i\sim0.1$--0.2, see Methods). We compare the excitation transfer behaviors of the two-mode systems with those of their single-mode counterparts, which have been experimentally realized on trapped-ion simulators in Refs.~\cite{so2024electrontransfer,gorman2018VAET,sun2024quantumsimulationspinbosonmodels} to various extents. In the following, we characterize the transfer dynamics by measuring the inverse lifetime of the donor population, $P_D=(\langle\sigma_z\rangle + 1)/2$, which is described by  \cite{skourtis1992photosynthesis, schlawin2021electrontransfer, so2024electrontransfer,padilla2025delocalizedexcitationtransferopen}:
\begin{equation}
    k_T=\frac{\int P_D(t)dt}{\int t P_D(t)dt}.
    \label{eq_lifetime_kT}
\end{equation} 
 This choice accounts for both the rate of dynamical equilibration and the steady-state population, which is analogous to the transfer efficiency, crucial for studying energy conversion in chemical processes.

{\bf Charge transfer (CT)} - We first study non-perturbative CT with $g_i\sim\omega_i$ in two different cases: degenerate ($\omega_1=\omega_2\equiv\omega$) and non-degenerate ($\omega_1>\omega_2$), where the approximate analysis breaks down.
In the degenerate case, we observe in Fig.~\ref{fig_CT}A that, in the two-mode model, the exothermic region characterized by monotonically increasing transfer rates with respect to the energy offset $\Delta E$, caused by both the broadening effect of strong electronic coupling to off-resonant states and by the dissipation rates that limit the transfer rates ($|V|>\gamma_i$) \cite{schlawin2021electrontransfer,so2024electrontransfer}, extends to a higher energy gap, $\Delta E \approx 4\omega$, compared to its single-mode counterpart ($\Delta E \approx 3\omega$). This extension can be attributed to the larger number of state configurations available when two vibrational modes (two dimensions) are present, rather than just one (one dimension). We note that it is not possible to adjust the Hamiltonian parameters of a single-mode model to reproduce this observation.
 
Moreover, we remark that resonant peaks at $\Delta E\approx\ell\omega$, with $\ell$ being an integer, appear when the energy difference is sufficiently high ($\Delta E \gtrsim 5\omega$). Under this energy gap condition, the initially localized donor state has significant overlaps with the eigenstates of the upper hybridized surface, leading to population trapping, as explained in Refs.~\cite{schlawin2021electrontransfer,so2024electrontransfer}. At these resonant peaks, the trapped population is released from the upper hybridized surface to the lower hybridized surface during the evolution. Owing to the large energy offset and the increased number of state configurations in the two-mode degenerate case (relative to the single-mode case), the steady states of these resonant transfers exhibit greater overlaps with the acceptor states, as shown in Figs.~\ref{fig_CT}B and \ref{fig_CT}C, where the final donor populations, $P_D(t_{\rm sim})$, are closer to zero than the steady state populations of the single-mode counterpart (see Fig.~\ref{fig_CT1M}B in Appendix \ref{SM_CT1M}). This results in enhanced transfer rates at large energy gaps $\Delta E\gtrsim 5\omega$, compared to the single-mode case.

Conversely, when $\omega_1>\omega_2$, the donor-acceptor energy landscape supports highly delocalized states along $y_2$, as the system lies deeply in the adiabatic regime ($|V|>\lambda_2/4$) \cite{schlawin2021electrontransfer, so2024electrontransfer}.
This introduces multiple states delocalized along the $y_2$ direction, in addition to the existing delocalization along $y_1$ provided by the fast vibrational mode levels. Together, these enable additional transfer channels among the highly delocalized two-dimensional vibronic states across the donor-acceptor energy gap. Thus, the transfer process in the non-degenerate case is less sensitive to the donor-acceptor energy offset, making it more robust to variations in this parameter. Due to our experimental resolution (see Appendix \ref{SM_numeric}), we cannot resolve the resonances associated with these transfer channels and instead observe a smooth transfer rate curve in Fig.~\ref{fig_CT}D. In addition, unlike the degenerate case, the transfer rates at the resonances with large $\Delta E$ are not evidently enhanced relative to those of the single-mode system. We emphasize that the non-degenerate two-mode spectrum observed in Fig.~\ref{fig_CT}D cannot be obtained by tuning the parameters of a single-mode model.

\begin{figure*}
    \centering
    \includegraphics[width=1\linewidth]{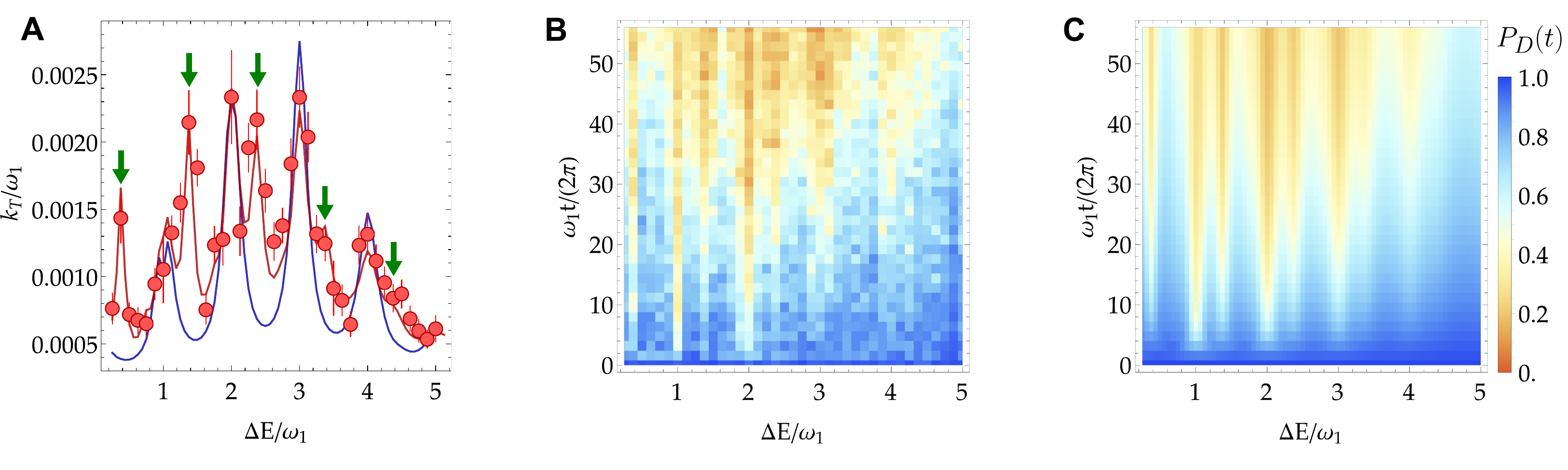}
    \caption{{\bf Transitioning from CT to VAET ($\omega_1\!>\!\omega_2$).} ({\bf A}) Transfer rate spectrum for $(V,g_{1},\gamma_{1},\omega_{2})=(0.063, 1.288, 0.015, 0.375)\omega_{1}$ and $(g_{2},\gamma_{2})=(0.607,0.067)\omega_2$. Red circles represent experimental data with error bars estimated via bootstrapping (see Methods). The solid red curve shows the transfer rate calculated from Eq.~\eqref{eq_master}, using the definition in Eq.~\eqref{eq_lifetime_kT} and including spin decoherence ($\gamma_{z}=0.0009 \omega_{1}$) and motional dephasing of both modes ($\gamma_{im}=0.0100\omega_{1}$, with $i=1,2$). The blue curve shows the numerical result for the single-mode case, where $\omega_2 = g_2 = \gamma_2 = \gamma_{2m} = 0$.
    ({\bf B and C}) Experimental and numerical donor population evolution $P_D(t)$ versus energy gap $\Delta E$ and the number of vibrational oscillations of the fast mode $\omega_{1} t/2\pi$, using the same parameters as in the red circles and solid red curve in {(A)}, respectively. Here, the detuning from the inner (outer) tilt mode, which encodes the fast (slow) vibrational energy, is set to $\delta_1=-2\pi\times8$ kHz ($\delta_2=-2\pi\times3$ kHz). Downward green arrows indicate nonadiabatic CT along $y_1$, assisted by single-phonon exchange with the slow mode via VAET.
    }
    \label{fig_transition}
\end{figure*}

{\bf Transitioning to VAET} - To verify that the additional transfer pathways---provided by the slow vibrational mode---are responsible for the smooth transfer profile shown in Fig.~\ref{fig_CT}D for the non-degenerate CT case, we decrease the electronic coupling strength and the vibronic coupling strength to the slow vibrational mode in Figs.~\ref{fig_transition}A-C (see Appendix \ref{SM_crossover} for more details on VAET-CT crossover). Choosing a weaker electronic coupling ($|V|<\lambda_1/4$) while maintaining strong vibronic coupling to the fast mode ($g_1>\omega_1$) places the system closer to the nonadiabatic CT regime along the fast mode direction. In this case, the donor and acceptor populations remain localized on their respective potential energy surfaces along $y_1$, and resonant excitation transfers occur between well-defined donor and acceptor vibronic states at $\Delta E \approx \ell_1\omega_1$, with $\ell_1$ being an integer  \cite{schlawin2021electrontransfer,so2024electrontransfer} (the one-dimensional case of Eq.~\eqref{eq_nonadiabatic}). In the single-mode scenario, where only $y_1$ is considered, this results in the manifestation of the vibrational mode structure in the transfer rate spectrum (solid blue curve in Fig.~\ref{fig_transition}A). Therefore, choosing a weaker electronic coupling allows us to distinguish the influences of the two vibrational degrees of freedom on the transfer dynamics when the second vibrational mode is involved. At the same time, lesser vibronic coupling to the slow mode reduces the distortion of the potential energy surfaces along $y_2$, making the transfer processes involving the slow mode to enter the VAET regime and introducing additional transfer rate resonances energetically enabled by the slow mode.

As shown by the resolved resonances in Fig.~\ref{fig_transition}A, the additional peaks around the nonadiabatic transfer resonances of the fast mode, which coincide with the single-mode spectrum (solid blue curve), are induced by the second mode and located at $\Delta E \approx \sqrt{(\ell_1\omega_1 + \ell_2\omega_2)^2-(2V)^2} \approx \ell_1 \omega_1 + \ell_2\omega_2$, where the $\ell_2\omega_2$ contribution comes from the VAET process bridging the energy gap between vibronic states defined by the fast mode (see the discussion of the VAET regime below). The processes associated with $\ell_2=1$ dominate and are more clearly observable because of the perturbative nature of VAET. With either a larger electronic coupling strength $V$ or a stronger vibronic coupling strength to the slow mode $g_2$, the additional resonances caused by the presence of the slow mode broaden and merge with the fast-mode transfer resonances into a smooth spectrum, as shown, for example, in Fig.~\ref{fig_CT}D. This highlights the crucial role of the simultaneous presence of fast and slow vibrational modes in reducing the CT dependence on the donor-acceptor energy gap through the additional transfer pathways.

%----------------------------
\begin{figure*}[t!]
    \centering
    \includegraphics[width=1\linewidth]{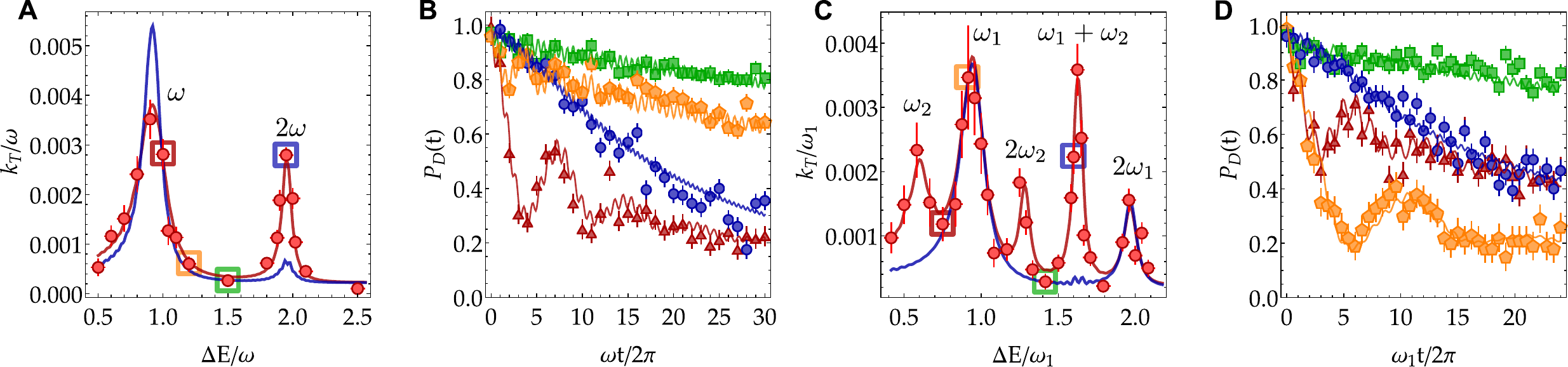}
    \caption{{\bf Transfer rates in the VAET regime.} ({\bf A}) Transfer rate spectrum of degenerate VAET ($\omega_1\!=\!\omega_2\!\equiv\!\omega$) with $(V,g_{1},\gamma_{1},g_{2},\gamma_{2})\!=\!(0.200, 0.200, 0.045, 0.220,0.018)\omega$. Red circles represent experimental data with error bars estimated via bootstrapping (see Methods). The solid red curve shows the transfer rate calculated from Eq.~\eqref{eq_master}, using the definition in Eq.~\eqref{eq_lifetime_kT} and including spin decoherence ($\gamma_{z}=0.0007 \omega$) and motional dephasing of both modes ($\gamma_{im}=0.0080\omega$, with $i=1,2$). The blue curve shows the numerical result of the single-mode VAET case, where $\omega_2\!=\!g_2\!=\!\gamma_2\!=\!\gamma_{2m}\!=\! 0$.
    ({\bf B}) Experimental and numerical donor population evolution $P_D(t)$ versus the number of vibrational oscillations $\omega t/2\pi$, using the same parameters as in the red circles and solid red curve in {(A)}, respectively. Here, the detunings from the two tilt modes for encoding the degenerate vibrational energies are both set to $\delta_1=\delta_2=-2\pi\times10$ kHz. ({\bf C and D}) The same layout as in {(A and B)} for the non-degenerate VAET case ($\omega_1\!>\!\omega_2$) with $(V,g_{1},\gamma_{1},\omega_{2},\gamma_{z},\gamma_{im})=(0.167, 0.275, 0.028, 0.667,0.0006,0.0068)\omega_{1}$ and $(g_{2},\gamma_{2})=(0.263,0.019)\omega_2$. The characteristic timescale is defined by the fast vibrational frequency $\omega_1$. In this case, the detuning from the inner (outer) tilt mode, which encodes the fast (slow) vibrational energy, is set to $\delta_1=-2\pi\times12$ kHz ($\delta_2=-2\pi\times8$ kHz). The red triangle, orange pentagon, green square, and blue circle markers correspond to $\Delta E=\{1.00,1.20,1.50,1.95\}\omega$ in the degenerate case and $\Delta E=\{0.75,0.92,1.42,1.60\}\omega_1$ in the non-degenerate case, respectively (marked by colored boxes in (A and C)). Near each resonant peak in the rate spectra, we label the total vibrational energy involved in the transfer.
    }
    \label{fig_VAETboth}
\end{figure*}

{\bf Vibrationally assisted exciton transfer (VAET)} - Unlike in the CT regime, where strong vibronic coupling distorts the donor-acceptor potential energy landscape that defines the vibronic eigenstates of the system, in the VAET regime ($g_i\!\ll\!\omega_i$), the vibrational modes are weakly coupled to the electronic degree of freedom and therefore only act as facilitators of the exothermic transfer, where units of vibrational energy are exchanged with the electronic sites during the excitation transfer at specific $\Delta E$ resonances, given by Eq.~\eqref{eq_VAETcondition} (see Fig.~\ref{fig_VAETboth}) \cite{li2021multimodeVAET}. Here, we also consider systems in the non-perturbative regime, characterized by strong electronic coupling ($|V| \sim \lambda_i/4$). For the vibrationally degenerate case ($\omega_1=\omega_2\equiv\omega$), shown in Fig.~\ref{fig_VAETboth}A, the transfer rate of the first resonance in the two-mode model is not enhanced compared to the single-mode case because only a single phonon from either degenerate mode can contribute to the transfer process at a time. In fact, the transfer rate becomes slightly slower due to the additional broadening from the dissipation of the second mode. On the contrary, at $\Delta E \approx \sqrt{(2\omega)^2-(2V)^2}$, VAET processes in which the total vibrational energy of $2\omega$ is supplied by a linear combination of the energy quanta from the two degenerate modes ($2\omega_1=2\omega_2=\omega_1+\omega_2$) interfere constructively. This leads to enhanced transfer rates compared to the single-mode case, where only the process involving a vibrational energy input of $2\omega_1$ can assist the transfer.

In the non-degenerate case $(\omega_1>\omega_2)$, the second-order processes occur at different donor-acceptor energy offsets, and their corresponding transfer resonances can be resolved experimentally (see Fig.~\ref{fig_VAETboth}C). As such, the presence of the slow mode does not affect the transfer rates at the resonances involving $\omega_1$ and $2\omega_1$ energy inputs. 
This is because, when we choose $\omega_1/\omega_2$ to be non-integer, there exists no non-trivial resonance involving an energy input of $\ell_2\omega_2$ from the slow mode, where $\ell_1\omega_1+\ell_2\omega_2=\ell\omega_1$.
However, additional resonances emerge instead from processes involving energy inputs given by linear combinations of the two non-degenerate vibrational energies. In Fig.~\ref{fig_VAETboth}C, we observe three additional resonances beyond the two existing resonances provided by the fast vibrational mode for the two-mode case. These peaks correspond to the processes involving vibrational energy inputs of $\omega_2$, $2\omega_2$, and $\omega_1+\omega_2$.
 
It is worth noting that, in both cases of mode degeneracy, the second-order processes associated with a vibrational energy contribution of $\omega_1 + \omega_2$ consist of two pathways (one phonon from the fast mode, then another from the slow mode, and vice versa) that interfere constructively (see Fig.~\ref{fig_concept}B). This coherent addition of the two rate amplitudes leads to an approximate two-fold enhancement of the peak at $\Delta E\approx1.63\omega_1$ in Fig.~\ref{fig_VAETboth}C with respect to the neighboring peaks at $\Delta E\approx1.29\omega_1$ and $\Delta E\approx1.97\omega_1$ that are associated with the second-order processes involving $2\omega_2$ and $2\omega_1$ energy inputs, respectively (see Appendix \ref{SM_coherence}). This observation suggests that approximately half of the enhancement at $\Delta E\approx 1.95\omega$ in the two-mode degenerate case with respect to the single-mode case is attributed to the $\omega_1+\omega_2$ pathways (see Appendix \ref{SM_coherence}). While constructive interference of the two vibrational modes requires full coherent control over both degrees of freedom, our results also demonstrate that this coherent enhancement remains resilient in the presence of dissipation.

In this work, we leverage the remarkable tunability of the trapped-ion platform to experimentally realize an open two-mode linear vibronic coupling model (LVCM) in two phenomenologically distinct regimes associated with charge transfer (CT) and vibrationally assisted exciton transfer (VAET), as well as their crossover. We simultaneously apply twelve carefully calibrated laser tones to independently control the coherent evolution of the qubit and the damping rates of two bosonic modes in a multi-species ion system. 
We observe enhanced transfer rates arising from the presence of the second mode across all vibronic coupling regimes---regardless of mode degeneracy---pointing out their differences and similarities. We also attribute these enhancements to coherent effects in the transfer pathways, which persist even in the presence of dissipation. Furthermore, our conclusions can be extended to two-site LVCM systems with more than two vibrational modes, where the same qualitative features identified in our findings remain (see Appendix \ref{SM_threemode}).
This observation also highlights the necessity of considering anharmonicity in quantum dynamics models, where quantum scrambling can occur at resonances, transitioning coherent quantum behavior into chaotic dynamics \cite{zhang2023scrambling}.

The experimental toolbox we deploy in this work is intrinsically scalable, as the same ion-laser couplings and sympathetic cooling techniques used to realize a two–mode, two–site LVCM can be extended to many vibrational modes and electronic sites without introducing additional physical overhead. In trapped-ion hardware, it is possible to include more molecular sites, as each extra qubit ion supplies a fully controllable two-level system that can encode a chromophore or charge-transfer center.
Each additional qubit ion or coolant ion also adds three collective bosonic modes, which can be directly used for reservoir engineering by tailoring the spectrum of sympathetic-cooling lasers, without the need for digitization \cite{macridin2018bosonicdigitization}. An arbitrary subset of these modes can then be endowed with individually programmable frequencies, coupling strengths, and dissipation rates, allowing the simulation of dissipative chemical dynamics in complex solvent environments \cite{plenio2013exciton,tiwari2013exciton}. Employing multiple engineered bosonic modes also enables the experimental realization of spin-boson models with tunable spectral densities, formed by a linear superposition of Lorentzian components \cite{leggett1987dissipative,lemmer2018engineering,sun2024quantumsimulationspinbosonmodels}, crucial for exploring phenomena related to non-Markovian dynamics \cite{debecker2024controlling}, such as coherence trapping \cite{huelga2012non,addis2014coherencetrapping, kamar2024dephasing, jiao2025protectingspinsqueezingdecoherence}, and dissipative quantum state engineering \cite{zhu2025steadystate}.

State-of-the-art trapped-ion quantum computing hardware already employs ion crystals with tens of ions while retaining individual ion control \cite{foss-feig2024, mueller2024quantumcomputinguniversalthermalization}. Therefore, scaling up the analog trapped-ion simulator presented here to a few tens of qubits and engineered bosonic modes is within reach. In the current experiment, for example, the number of qubits and engineered bosonic modes is limited by purely technical factors, such as the ion chain vacuum lifetime and the available laser power. Trapped-ion systems also provide tunable long-range spin-spin couplings and high-fidelity entangled state generation via Molmer-Sorensen interactions \cite{monroe2021programmable}, which can be used to mimic long-range electronic couplings in Frenkel-exciton models \cite{Jang2018} and to study the role of delocalization in exciton transfer \cite{fassioli2014vibration, sneyd2021delocalization, padilla2025delocalizedexcitationtransferopen}.

This work establishes a clear, hardware-efficient roadmap for scalable trapped-ion analog platforms to investigate a wide range of open-system spin-boson models with multiple electronic configurations and vibrational modes, paving the way for the simulation of singlet fission processes \cite{collins2023fission,campaioli2024optimisationultrafastsingletfission}, electron-phonon propagation in condensed matter physics \cite{knozer2022holstein}, and realistic photochemical and bioenergetic processes \cite{sneyd2021delocalization}. In particular, we show how trapped ions enable the simulation of these models in the intermediate coupling regime, with the reorganization energy and electronic coupling strength being of the same order, which can be challenging for existing classical methods \cite{somoza2019dissipation,kang2024chemical,fassioli2014vibration}. 

%----------------------------
% materials and methods
%----------------------------
\section*{Methods}
\label{sec_method}
%----------------------------
% experimental setup
%----------------------------
\subsection*{Experimental setup}

The experimental system used for this study is thoroughly described in Ref.~\cite{so2024electrontransfer}, where the dynamics of the single-mode LVCM in the CT regime are realized with four Raman 355 nm and two 435.5 nm laser tones. In this work, we include two additional tones to each beam to generate the terms associated with the second vibrational mode, resulting in a total of ten laser tones on the 355 nm and 435.5 nm lasers, over which we have full control of both amplitudes and frequencies. We also use the collective modes of the two-ion chain along both radial directions of the trap---specifically the $y$ and $z$ tilt modes ($\omega_{\text{tilt},y}\equiv\omega_{\text{tilt},1}=2\pi\times3.151$ MHz and $\omega_{\text{tilt},z}\equiv\omega_{\text{tilt},2}=2\pi\times3.740$ MHz)---to encode the vibrational degrees of freedom in the two-mode LVCM Hamiltonian. The unused radial collective modes are the center-of-mass modes with $\omega_{\text{com},y}\equiv\omega_{\text{com},1}=2\pi\times3.318$ MHz and $\omega_{\text{com},z}\equiv\omega_{\text{com},2}=2\pi\times3.882$ MHz. Since the frequency separations among the available radial collective modes are set to be sufficiently large ($\Delta\omega_{\text{trap},j}\gtrsim2\pi\times140$ kHz), undesired off-resonant spin-phonon interactions, which are proportional to $g_j/(\mu_i-\omega_{\text{trap},j})$, can be neglected \cite{pagano2025varenna,monroe2021programmable}. Here, $\mu_i$ is the 355 nm laser frequency detuning from the qubit resonance, used to generate the spin-dependent force on the target radial tilt mode $i$, and $\omega_{\text{trap},j}$ is the relevant collective radial frequency considered for the off-resonant spin-phonon drive ($j\neq i$).

\subsection*{Experimental sequence}

% experimental sequence
The experimental sequence (see Fig.~\ref{fig_experiment}B) begins with Doppler cooling and Raman-resolved sideband cooling on all four collective radial modes of the chain, which results in an initial phonon population of both radial tilt modes, $\bar{n}_{0,i} \sim 0.1$--0.2, which is set to match the independently measured $\bar{n}_i$. We then apply a $\pi/2$ pulse to map the $z$ qubit basis to the $y$ basis and two consecutive displacement operations via the spin-dependent optical force to prepare the system in the donor vibronic state, $\ket{D}\bra{D}\otimes \rho_{1-}\otimes \rho_{2-}$, where $\rho_{i-}=\sum_{n_i} e^{- n_i\omega_i/k_B T_i}\ket{n_{i-}}\bra{n_{i-}}$ represents a thermal state with temperature $k_B T_i\approx \omega_i/\log(1 + 1/\bar{n}_i)$, and $\ket{n_{i\pm}}=\mathcal{D}(\pm g_i/2\omega_i)\ket{n_i}$ are displaced Fock states associated with vibrational mode $i$. For simulating the open-system LVCM dynamics described by Eq.~\eqref{eq_master}, we simultaneously apply the six 355 nm, four 435.5 nm, and two 935 nm laser tones. After evolving the system for time $t_{\rm sim}$, we apply another $\pi/2$ pulse to map the quantum state in the $y$ basis back to the $z$ qubit basis and measure the probability of the system being in the donor state $P_D=(\langle\sigma_z\rangle + 1)/2$ via state-dependent fluorescence.

%----------------------------
% transfer rate data analysis
%----------------------------
\subsection*{Transfer rate data analysis}

% introduce modified transfer rate formula
We experimentally simulate the LVCM dynamics from $t=0$ ms to $t=t_{\rm sim}$, where the finite simulation time $t_{\rm sim}$ ranges from 2 to 8 ms, corresponding to 24--65 vibrational cycles of the fast mode (defined by $\omega_1t/2\pi$), within which the system undergoing the first transfer resonance reaches equilibrium. Since Eq.~\eqref{eq_lifetime_kT} defines the transfer rate in the limit of $t_{\rm sim}\rightarrow\infty$, an offset correction is required when applying it to finite-time dynamics. Particularly, in the case of a strictly localized donor population, $P_D(t)= {\rm constant}$, Eq.~\eqref{eq_lifetime_kT} still yields a nonzero transfer rate of $k_T=\frac{2}{t_{\rm sim}}$, which approaches zero only as $t_{\rm sim}\rightarrow\infty$. Therefore, we need to remove the undesired background from the finite-time evaluation of the non-zero steady-state donor population in the transfer rate calculations, as follows \cite{so2024electrontransfer,padilla2025delocalizedexcitationtransferopen}:
\begin{equation}
    k_T=\frac{\int_{0}^{t_\text{sim}} P_D(t)dt}{\int_{0}^{t_\text{sim}} t P_D(t)dt} - \frac{2}{t_\text{sim}}.
    \label{eq_mod_lifetime_kT}
\end{equation}
We note that this background correction does not alter the characteristic features of the transfer rate spectra. By interpolating the donor population probability $P_D(t)$ for both the experimental and numerical data using the same time steps and applying the modified formula above, we obtain the transfer rates of the dynamics reported in the main text. As in Ref.~\cite{so2024electrontransfer}, we also use a resampling (bootstrapping) procedure at each time step of the $P_D(t)$ measurements to estimate the uncertainties of the extracted transfer rates. For each time step, we treat the experimental error as the standard deviation of a normal distribution centered at the measured mean value, from which the resampled datasets are drawn. We then use the standard deviation of the transfer rates calculated from the resampled datasets as the uncertainty associated with the reported rates.

%----------------------------
% acknowledgements
%----------------------------
\begin{acknowledgments}
We acknowledge Diego Fallas Padilla for careful reading of the manuscript and helpful suggestions.
This work was supported in part by the NOTS cluster operated by Rice University's Center for Research Computing (CRC).
%----------------------------
% funding
%----------------------------
G.P. acknowledges support from the Welch Foundation Award (grant no. C-2154), the Office of Naval Research Young Investigator Program (grant no. N00014-22-1-2282), the NSF CAREER Award (grant no. PHY-2144910), the Army Research Office (W911NF22C0012), and the Office of Naval Research (grant no. N00014-23-1-2665). We acknowledge that this material is based on work supported by the U.S Department of Energy, Office of Science, Office of Nuclear Physics under the Early Career Award (grant no. DE-SC0023806). The isotopes used in this research were supplied by the US Department of Energy Isotope Program, managed by the Office of Isotope R$\&$D and Production. H.P. acknowledges support from the NSF (grant no. PHY-2513089) and the Welch Foundation (grant no. C-1669). Work at the Center for Theoretical Biological Physics was supported by the NSF (grant no. PHY-2019745). J.N.O. was also supported by the NSF (grant no. PHY-2210291). P.G.W. was also supported by the D. R. Bullard-Welch Chair at Rice University (grant no. C-0016).

\vspace{1 em}
\noindent\textbf{Author contributions:} V.S., M.D.S., A.M., G.T., R.Z., and G.P. contributed to the experimental design, construction, data collection, and analysis of this experiment. V.S., M.Z., H.P., J.N.O., P.G.W., and G.P. contributed to the paper’s conceptualization and supporting theory and numerics. All authors contributed to the writing and revision of the manuscript.

\vspace{1 em}
\noindent\textbf{Competing interests:} R.Z. is a cofounder and chief executive officer of TAMOS Inc. G.P. is a cofounder and chief scientist of TAMOS Inc. The other authors declare no competing interests.

\vspace{1 em}
\noindent\textbf{Data availability:} The authors declare that the data supporting the findings of this study are available within the paper, its appendices, and upon request.

\end{acknowledgments}

%----------------------------
% references
%----------------------------
\bibliography{refs}% Produces the bibliography via BibTeX.

%----------------------------
% supplementary materials
%----------------------------

\appendix
% SM setup
\setcounter{figure}{0}
\renewcommand{\figurename}{Figure}
\renewcommand{\thefigure}{S\arabic{figure}}

%----------------------------

%----------------------------
% system calibration
%----------------------------
\section{System calibration} \label{SM_calibration}

As described in the main text, we have full control over the system parameters and the properties of the reservoirs. Prior to realizing the two-mode LVCM dynamics with the twelve laser tones, we independently calibrate each term of the Hamiltonian and the motional cooling of each vibrational mode using their corresponding laser configurations, following the procedure detailed in  Ref.~\cite{so2024electrontransfer}, with additional steps for the drives associated with the second vibrational mode. These additional steps include calibrations of $g_2$, $\omega_2$, $\gamma_2$, $\bar{n}_2$, and $\bar{n}_{0,2}$, while maintaining the pre-calibrated values of $\Delta  E$, $V$, $g_1$, $\omega_1$, $\gamma_1$, $\bar{n}_1$, and $\bar{n}_{0,1}$ (relevant to single-mode systems). The following table lists the values of the system parameters and motional cooling rates used to realize the LVCM dynamics in the main text:
%-----------------------------
\begin{table}[h!]
\centering
\begin{tabular}{|c|c|c|c|c|c|c|c|c|}
\hline
\textbf{ Reported } & \multicolumn{8}{c|}{\textbf{ Trapped-ion parameters [$2\pi \times$ kHz] }} \\
\cline{2-9}
 \textbf{data} & $\Delta E$ & $V$ & $\omega_1$ & $g_1$ & $\gamma_1$ & $\omega_2$ & $g_2$ & $\gamma_2$ \\
\hline
Figs.~\ref{fig_CT}A-B & $\;2\text{--}28\;$ & $\;1\;$ & $\;5\;$ & $\;6\;$ & $\;0.18\;$ & $\;5\;$ & $\;5.5\;$ & $0.2$ \\
\hline
Figs.~\ref{fig_CT}C-D & $\;3\text{--}40\;$ & $\;1.1\;$ & $\;8\;$ & $\;8.23\;$ & $\;0.18\;$ & $\;3\;$ & $\;2.98\;$ & $0.08$ \\
\hline
Figs.~\ref{fig_transition}A-B & $\;2\text{--}40\;$ & $\;0.5\;$ & $\;8\;$ & $\;10.3\;$ & $\;0.12\;$ & $\;3\;$ & $\;1.82\;$ & $0.2$ \\
\hline
Figs.~\ref{fig_VAETboth}A-B & $\;5\text{--}25\;$ & $\;2\;$ & $\;10\;$ & $\;2\;$ & $\;0.45\;$ & $\;10\;$ & $\;2.2\;$ & $0.18$ \\
\hline
Figs.~\ref{fig_VAETboth}C-D & $\;5\text{--}25\;$ & $\;2\;$ & $\;12\;$ & $\;3.3\;$ & $\;0.34\;$ & $\;8\;$ & $\;2.1\;$ & $0.15$ \\
\hline
\end{tabular}
\caption{Trapped-ion interaction settings used for simulating the LVCM dynamics in the main text. The values of the system parameters ($\Delta E,\;V,\;\omega_1,\;g_1,\;\omega_2,\;g_2$) and vibrational dissipation rates ($\gamma_1,\;\gamma_2$) are determined by the frequency and power of the laser tones used to generate the associated ion-light interactions.}
\label{table:exprange}
\end{table}

As shown in Table \ref{table:exprange}, in this work, we demonstrate precise control and wide-range tunability of ion-light interactions on our quantum simulator, which allow us to explore the rich dynamics of multi-mode LVCM systems in different vibronic coupling regimes and vibrational mode degeneracy.

%----------------------------
% numerical simulations
%----------------------------
\section{Numerical simulations} \label{SM_numeric}

% describe the numerical simulations
We use a QuTiP-based Python package \cite{johansson2013qutip} to numerically obtain the LVCM dynamics for theoretical investigations and to compare with the experimental data. Due to experimental imperfections, we include additional decoherence processes to our simulations of Eq.~\eqref{eq_master}, as follows:
\begin{eqnarray}
    \frac{\partial\rho}{\partial t}&=&-i[H,\rho] + \sum_{i=1,2}\{\gamma_i (\bar{n}_i+1)\mathcal{L}_{a_i}[\rho] + \gamma_i \bar{n}_i \mathcal{L}_{a^\dagger_i}[\rho]\} \nonumber \\ 
    &\quad& \quad+\;\gamma_z\mathcal{L}_{\sigma_y}[\rho]+ \sum_{i=1,2}\gamma_{im}\mathcal{L}_{c_{im}}[\rho],
    \label{eq_master_mod}
\end{eqnarray}
where the jump operator $\sigma_y$ and its corresponding rate $\gamma_z$ account for spin dephasing induced by laser power fluctuations in the rotated spin basis ($z \leftrightarrow y$), while the jump operators $c_{im}= a_i^\dagger a_i$ and their corresponding rates $\gamma_{im}$ consider the motional dephasing of the radial tilt modes due to trap frequency fluctuations \cite{fluhmann2019encoding, so2024electrontransfer}. From the comparison between the numerical calculations and experimental data, we obtain $\gamma_z/2\pi = 7$ Hz and $\gamma_{im}/2\pi = 80$ Hz for all cases. These experimental imperfections reduce the resolution in $\Delta E$ and suppress the sharp (and narrow) peaks in the transfer rate spectra.

%----------------------------
%Weak-coupling regime
%----------------------------
\section{Perturbative analysis} \label{SM_weakcoupling}

\subsection{Weak electronic coupling}
\label{SM_weakcouplingET}

% introduce FGR
Throughout the main text, we focus our investigation of the two-mode CT process in the strong electronic coupling regime ($|V|\sim\lambda_i/4$), where there is no analytical description for the rate of the transfer dynamics. However, in the weak electronic coupling regime, where $|V|\ll\lambda_i/4$, the eigenstates of the two-mode LVCM can be approximated to the two-dimensional uncoupled donor and acceptor vibronic states, where we treat the electronic coupling term $V\sigma_x$ as a perturbation to the uncoupled system, described by $H - V\sigma_x$ with $g_i\gtrsim\omega_i$. In this case, the transfer rates are given by the Fermi's golden rule (FGR) for the transitions between the donor and acceptor vibronic states, as follows \cite{leggett1987dissipative, skourtis1992photosynthesis, schlawin2021electrontransfer, onuchic1987twomode, so2024electrontransfer}:
\begin{equation}
    k_T\!=\!2\pi|V|^2\!\sum_{\substack{n_{1-},n_{1+}\\n_{2-},n_{2+}}}p_{n_{1-}}p_{n_{2-}}\text{FC}_{\substack{n_{1-},n_{1+}\\n_{2-},n_{2+}}} {\rm L}(E_{DA},\gamma_{\rm eff}),
    \label{eq_kFermi}
\end{equation}
where ${\rm FC}_{\substack{n_{1-},n_{1+}\\n_{2-},n_{2+}}}=|\langle n_{1-} \ket{n_{1+}}\langle n_{2-} \ket{n_{2+}}|^2$ is the convolution of the Franck-Condon factors of the two vibrational modes, which describes the total two-dimensional overlap between the displaced Fock wavefunctions, and $p_{n_{i-}}$ is the initial phonon population of the vibrational mode $i$ in the donor state. Here, we account for the dissipation of the vibrational modes by including a Lorentzian energy distribution to the resonances at $E_{DA}\equiv E_{D,n_{1-},n_{2-}}-E_{A,n_{1+},n_{2+}}=\Delta E$, the difference between the eigenenergies of the uncoupled donor $\ket{D}\otimes\ket{n_{1-}}\otimes\ket{n_{2-}}$ and acceptor $\ket{A}\otimes\ket{n_{1+}}\otimes\ket{n_{2+}}$ vibronic states. The line-broadening profile takes the form:
\begin{equation}
    {\rm L}(E_0,\gamma)=\frac{\gamma/2\pi}{E_0^2 + \gamma^2/4}.
\end{equation}
The effective spectral width of the broadening is $\gamma_{\rm eff} = C_1\gamma_1 + C_2\gamma_2$, where the vibrational mode $i$ is independently subjected to the dissipation rate $\gamma_i$ with $\gamma_i \gtrsim |V|$, and $C_i$ corresponds to its correction factor, which is explained below. The effective broadening arises from the convolution of two Lorentzian distributions of widths $C_1\gamma_1$ and $C_2\gamma_2$. This description can be extended to the perturbative CT systems with a higher number of vibrational modes ($i>2$).

\begin{figure}[t!]
    \centering
    \includegraphics[width=1\linewidth]{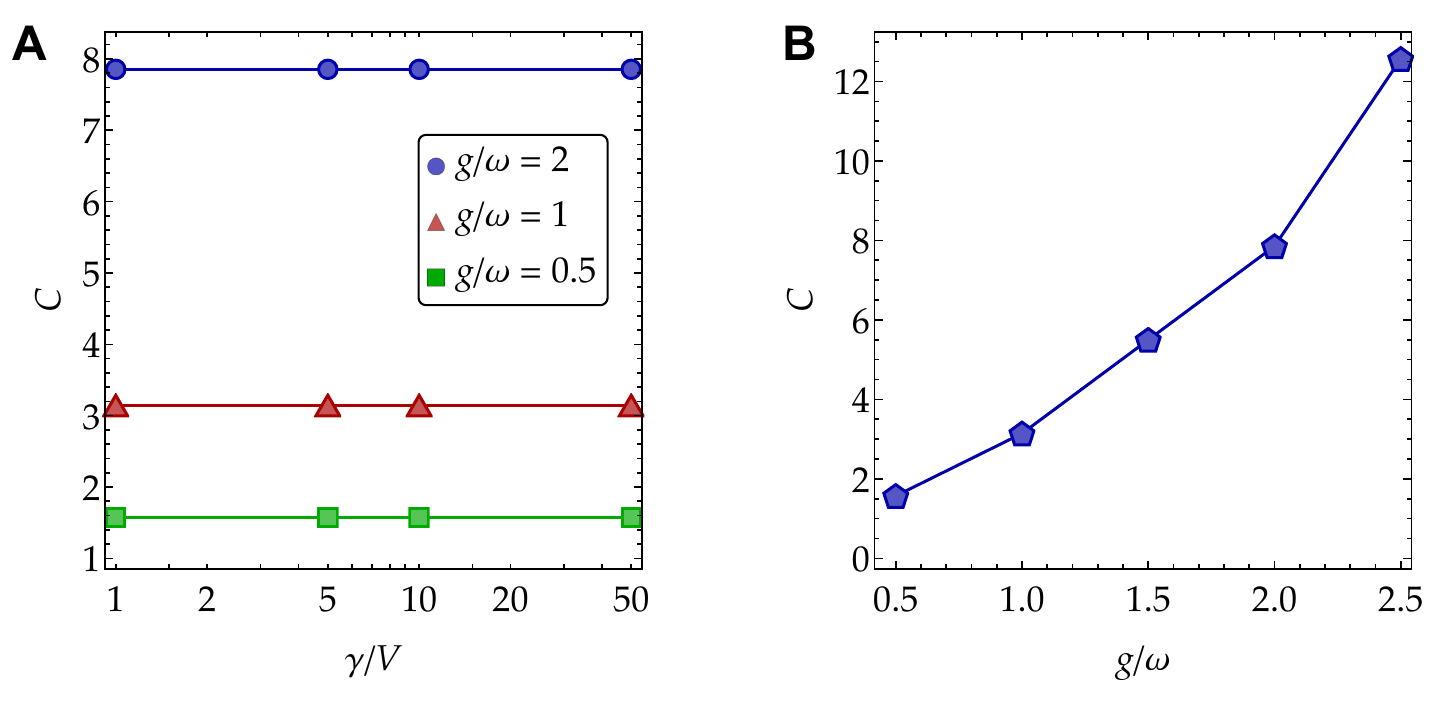}
    \caption{{\bf Line-broadening corrections in the weak electronic coupling regime.} Empirically determined width correction factor $C$ (connecting point markers) for the single-mode CT process with $V = 0.005\omega$ and $\bar{n}=\bar{n}_0=0.01$, plotted as a function of ({\bf A}) $\gamma/V$ and ({\bf B}) $g/\omega$.
    }
    \label{fig_broadeningET}
\end{figure}

Regarding the width of the line broadening associated with the vibrational dissipation in perturbative CT systems, we numerically find that it depends not only on the dissipation rate $\gamma_i$ but also on the displacement given by $g_i/\omega_i$. This dependence is related to the fact that the Lindbladian eigenvalues are diagonal in the non-displaced Fock basis, and not in the eigenbasis of the vibronic system, which causes the collapse operator to act on both the donor and acceptor vibronic states \cite{zhu2025steadystate}.
To demonstrate this finding, we revisit the single-mode CT case with a vibrational energy $\omega$ and a vibronic coupling $g$ subject to vibrational dissipation at a rate $\gamma$. The effective full width at half maximum of the Lorentzian broadening is presumed to be $\gamma_{\rm eff}= C\gamma$. As shown in Fig.~\ref{fig_broadeningET}A, we empirically estimate a fixed value of the correction factor $C>1$ for a given $g/\omega$ by comparing the transfer rate spectra obtained from the master equation with those from the FGR. Since $C$ is found to be independent of $\gamma$, we can deduce that the effective broadening width $\gamma_{\rm eff}$ is linearly proportional to $\gamma$. By varying $g/\omega$ for fixed $\gamma$ and $V$ in Fig.~\ref{fig_broadeningET}B, we observe that $C$ grows as $\sim (g/\omega)^2$. For instance, when $g/\omega = 2.5$, we get $C \approx 4\pi$, the correction factor used in Ref.~\cite{schlawin2021electrontransfer}. Although $C$ seems to approach the value of 1 as $g/\omega$ goes to 0, this limit is forbidden by the perturbation criterion of the regime, which requires $|V|\ll\lambda/4$. However, $C=1$ is appropriate for the analysis of the perturbative VAET dynamics in Appendix \ref{SM_weakcouplingVAET}, where $g$ is the perturbation to the uncoupled vibronic system, whose eigenstates are the products of the electronic states and the non-displaced Fock states.

It is also worth noting that we have used the definition in Eq.~\eqref{eq_lifetime_kT} for the transfer rate to capture both the time it takes for the system to equilibrate and the steady-state population of the dynamics throughout the main text for our non-perturbative studies. As pointed out in Refs.~\cite{schlawin2021electrontransfer,padilla2025delocalizedexcitationtransferopen}, given an exponentially decaying dynamics with no remaining population in the initial state, the transfer rate calculated by Eq.~\eqref{eq_lifetime_kT} converges to the inverse of the time constant at sufficiently large $t_{\rm sim}$. However, when the population transfer is not complete ($P_D^{\rm SS}\equiv P_D(t\rightarrow\infty)>0$), the absolute values of the rates extracted from Eq.~$\eqref{eq_lifetime_kT}$ differ from those obtained via the exponential fits despite retaining the same qualitative behaviors in the transfer rate spectra. Therefore, when comparing the transfer rates of the perturbative dynamics given by the master equation with the FGR predictions, it is more accurate to use the inverse time constants from the exponential fits rather than the rates obtained from Eq.~$\eqref{eq_lifetime_kT}$.

%----------------------------
\begin{figure}[t!]
    \centering
    \includegraphics[width=1\linewidth]{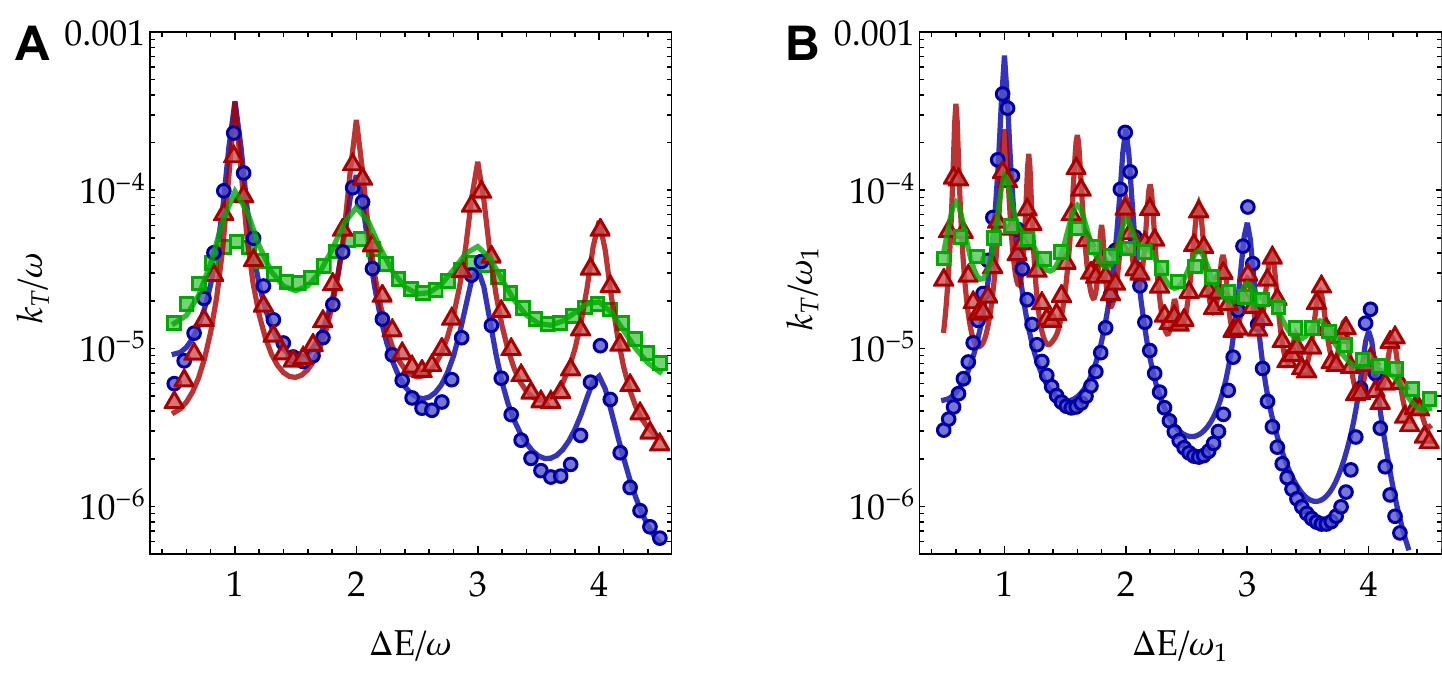}
    \caption{{\bf Transfer rates in the weak electronic coupling regime.} ({\bf A}) Transfer rate spectra for the degenerate two-mode CT case ($\omega_1\!=\!\omega_2\!\equiv\!\omega$) with $(V,g_{1},g_{2})=(0.005, 1, 1)\omega$ and $\bar{n}_{0,i}=\bar{n}_{i}=0.01$. Red and green data correspond to $(\gamma_1,\gamma_2)=(0.04, 0.01)\omega$ and $(0.05, 0.15)\omega$, respectively. Blue data are the numerical results of the single-mode CT process, where $\omega_2\!=\!g_2\!=\!\gamma_2\!=\!0$ and $\gamma_1\!\equiv\!\gamma\!=\!0.05\omega$.
    ({\bf B}) Transfer rate spectra for the non-degenerate two-mode CT case ($\omega_1 > \omega_2$) with $(V,g_{1},g_{2},\omega_{2})=(0.005, 1, 1, 0.6)\omega_{1}$. Similarly, red and green data correspond to $(\gamma_1,\gamma_2)=(0.025, 0.005)\omega_1$ and $(0.025, 0.075)\omega_1$, respectively. Blue data are the numerical results of the single-mode CT process, where $\omega_2 = g_2 = \gamma_2 = 0$ and $\gamma_1 \equiv \gamma = 0.025\omega_1$. The solid curves show the transfer rates obtained from the exponential fits of the master equation dynamics, while the point markers are their corresponding FGR predictions from Eq.~\eqref{eq_kFermi}. 
    }
    \label{fig_weakcoupling}
\end{figure}
%----------------------------

% discuss overall fig
By considering these effects in our analysis, we observe that the FGR calculations agree well with the numerical results of the master equation in Eq.~\eqref{eq_master} for weak electronic couplings ($V = 0.02 \times \lambda_i/4$), as shown in Fig. \ref{fig_weakcoupling}. Despite the overall increase in the energetically allowed transfer robustness, there are some subtle differences in the transfer behaviors between the degenerate ($\omega_1 = \omega_2 \equiv \omega$) and non-degenerate ($\omega_1 > \omega_2$) cases when compared to their single-mode ($\omega_2 = g_2 = \gamma_2 = 0$) counterparts. Thus, we shall discuss their features separately:

% discuss symmetric case
{\bf Degenerate case $(\omega_1=\omega_2\equiv\omega)$} - As shown by the red and blue curves in Fig.~\ref{fig_weakcoupling}A, the transfer rates of the two-mode system are lower to those of the single-mode system around $\Delta E = \omega$, and they become larger at higher resonances ($\Delta E = m\omega, m\in\mathbb{N},m>1$) for $\gamma_1+\gamma_2=\gamma$. This can be understood by comparing the FGR formula of the single-mode and two-mode cases.  At a low temperature ($\bar{n}_i \sim 0.01$), the initial population in both cases dominantly occupies the ground level of the donor well. For $\Delta E = m\omega, m\in\mathbb{N}$, the transfer rates are proportional to the wavefunction overlaps between the donor and acceptor states, given by ${\rm FC}_{\text{2M}}\approx \sum_{k_1=0}^m\sum_{k_2=0}^m\delta_{k_1+k_2,m}|\langle 0_{-} \ket{k_{1+}}|^2|\langle 0_{-} \ket{k_{2+}}|^2$ in the degenerate two-mode case and ${\rm FC}_{\text{1M}}\approx |\langle 0_{-} \ket{m_{+}}|^2$ in the single-mode case, where we suppose $p_{n_{i-}}\equiv p_{n_{-}}\approx p_{0_{-}}\approx1$ and $|\langle n_{i-} \ket{n_{i+}}|\equiv|\langle n_{-} \ket{n_{+}}|\leq1$ for simplicity. Intuitively, the vibrational modes of the two-dimensional donor well can ``share'' the energy difference, increasing the likelihood of transitions to higher excited states of the two-dimensional acceptor well (for $m>1$) despite the lessened individual couplings from the smaller wavefunction overlaps as compared to the single-mode case. For example, for a well separation of $g_i/\omega_i = 1$ between the donor and acceptor energy surfaces, the Franck-Condon factors of the two systems decrease with increasing $\Delta E = m\omega$ as follows: ${\rm FC}_{\text{2M}}=\{0.271, 0.271, 0.180, 0.124\}$ and ${\rm FC}_{\text{1M}}=\{0.368, 0.184, 0.061, 0.015\}$ for $m=\{1,2,3,4\}$, which explains the comparative features of the red and blue data points in Fig.~\ref{fig_weakcoupling}A.

However, for increasing width $\gamma_{\rm eff}$, the transfer dependence on $\Delta E$ can be diminished at the expense of the transfer rates at resonances. As shown in the green curve of Fig.~\ref{fig_weakcoupling}A, when we increase the dissipation rate on the slow mode, the widths of the two-mode transfer resonances grow despite the lowered peak values, which increases the transition probabilities for off-resonant processes, thus making excitation transfer more robust to the energy offset in the donor-acceptor system.

% discuss asymmetric case
{\bf Non-degenerate case $(\omega_1>\omega_2)$} - The additional resonances, provided by the slow mode, can also increase the transfer robustness to $\Delta E$ (see Fig.~\ref{fig_weakcoupling}B). Extra broadening of the resonances from increased $\gamma_{\rm eff}$ can further enhance this robustness. Meanwhile, the transfer rates at resonances associated with the fast mode ($\Delta E = m\omega_1, m\in\mathbb{N}$) are evidently reduced compared to the single-mode system, differently from the degenerate case. This decrease can be explained by the significantly lowered overlaps of the wavefunctions $\left(\sum_{\substack{n_{1-},n_{1+}\\n_{2-},n_{2+}}}{\rm FC}_{\substack{n_{1-},n_{1+}\\n_{2-},n_{2+}}}\ll\sum_{\substack{n_{-},n_{+}}}{\rm FC}_{\substack{n_{-},n_{+}}}\right)$ due to the mismatch of the resonances between those associated with the fast vibrational mode and those provided by the slow vibrational mode ($\Delta E = m\omega_1 \neq m\omega_2$).

\subsection{Weak vibronic coupling (VAET)}
\label{SM_weakcouplingVAET}

To gain insights into the VAET regime, we shall employ a similar perturbative analysis of the two-mode model in the weak vibronic coupling regime ($g_j\ll \omega_j$) \cite{gorman2018VAET, li2021multimodeVAET}. The unperturbed system is now the uncoupled, non-displaced vibronic system, described by:
\begin{equation}
    H_{\rm 2M,unc}^{\rm VAET} = \frac{\Delta E}{2}\sigma_z + V\sigma_x +\sum_{j=1}^2 \omega_j a^\dagger_j a_j.
\end{equation}
The eigenstates of this system are given by:
\begin{eqnarray}
    \ket{e_\pm,\!n_1,\!n_2}\!=\! \Biggl(\!&\pm&\!\frac{\epsilon\pm\Delta E/2}{\sqrt{2\epsilon(\epsilon\!\pm\!\Delta E/2)}}\!\ket{\uparrow}\!+\!\frac{V}{\sqrt{2\epsilon(\epsilon\!\pm\!\Delta E/2)}}\!\ket{\downarrow}\!\Biggr) \nonumber \\
    &\otimes& \ket{n_1} \otimes \ket{n_2},
    \label{eq_VAETpertsystem}
\end{eqnarray}
which correspond to the eigenenergies $E_{e_\pm,n_{1},n_{2}} = \pm\epsilon + n_1\omega_1 + n_2\omega_2$, where $\epsilon\equiv\sqrt{\left(\frac{\Delta E}{2}\right)^2+V^2}$, and $n_j\in \mathbb{N}$. Without the vibrational displacements being parts of the uncoupled vibronic system, the concept of state-dependent potential energy landscape does not apply here, and $H_{\rm Int}^{\rm VAET}=\sum_{j=1}^2\frac{g_j}{2}\sigma_z(a_j+a_j^\dagger)$ rather acts as a perturbation that induces transitions between the states with ladder-like energy levels. By transforming into the eigenstate basis defined in Eq.~\eqref{eq_VAETpertsystem}, the vibronic perturbation can be written as:
\begin{eqnarray}
    H_{{\rm Int},e}^{\rm VAET} &=& \sum_{j=1}^2 \frac{g_j}{2\epsilon}\left(\frac{\Delta E}{2}\tilde{\sigma}_z-V\tilde{\sigma}_x\right)(a_j+a_j^\dagger),
\end{eqnarray}
where $\tilde{\sigma}_{x,z}$ are the Pauli operators in the eigenstate basis. From the first-order perturbation theory, single-phonon exchange processes are allowed, and the transfer rates are given by the first-order transition probability amplitude $C_T^{(1)}$ and the final density of states $\rho_F(E_I)$:
\begin{eqnarray}
    k_{T}^{(1)} &=& 2\pi\left|C_T^{(1)}\right|^2\rho_F(E_I)\nonumber\\
    &=&2\pi\left|\frac{V}{2\epsilon}\right|^2 \sum_{\substack{n_{1I},n_{1F}\\n_{2I},n_{2F}}} p_{n_{1I}}p_{n_{2I}}\sum_{j=1}^2|g_j|^2\delta_{n_{kF},n_{kI}}
    \label{eq_kFermiVAET1}\\&\quad&\times\left[n_{jI}\delta_{n_{jF},n_{jI}-1}\!+\!(n_{jI}+1)\delta_{n_{jF},n_{jI}+1}\right]{\rm L}(E_{IF},\gamma_j), \nonumber
\end{eqnarray}
where index $k\neq j$, specifically $(j,k)$ can only be either $(1,2)$ or $(2,1)$, and
$E_{IF} \equiv E_{e_+,n_{1I},n_{2I}}-E_{e_-,n_{1F},n_{2F}}=2\epsilon + (n_{1I}-n_{1F})\omega_1 + (n_{2I}-n_{2F})\omega_2$ is the energy difference between the initial $\ket{e_+,n_{1I},n_{2I}}$ and final $\ket{e_-,n_{1F},n_{2F}}$ eigenstates. The allowed transfers in Eq.~\eqref{eq_kFermiVAET1} describe the processes in which the electronically coupled system exchanges single-phonon energy with a vibrational mode to enable the excitation transfer \cite{gorman2018VAET,sun2024quantumsimulationspinbosonmodels}.

\begin{figure}[t!]
    \centering
    \includegraphics[width=1\linewidth]{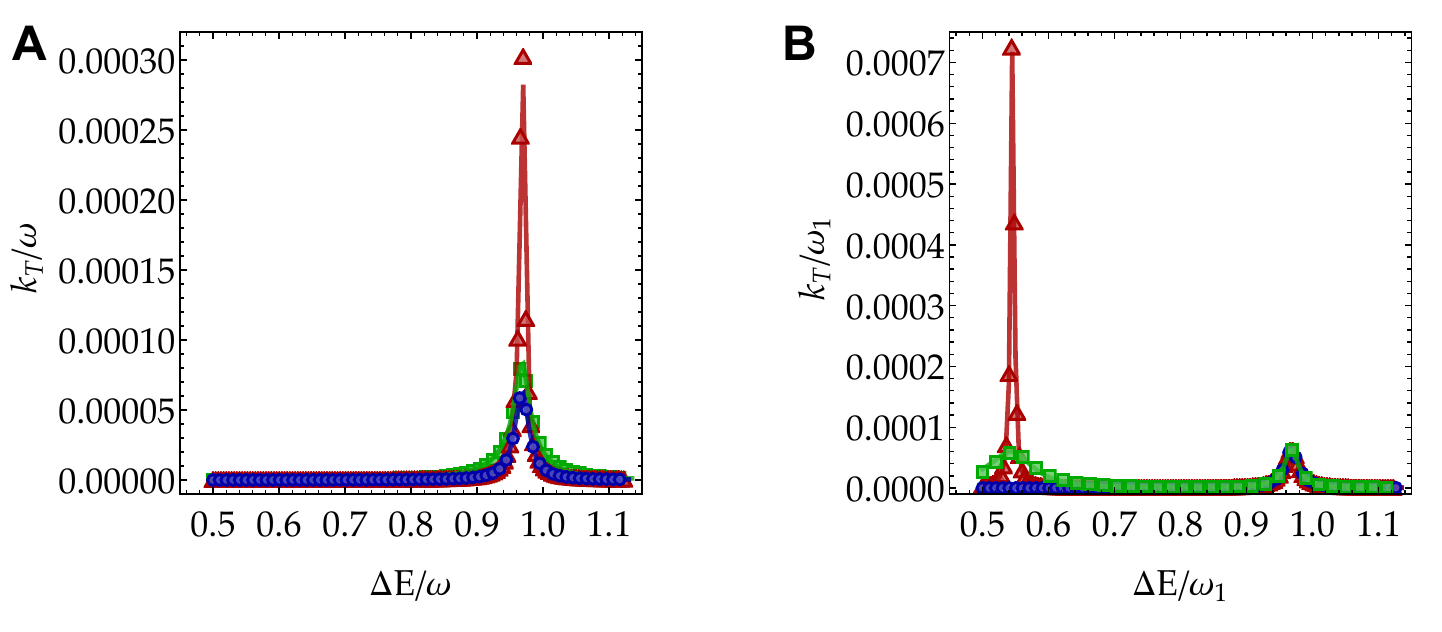}
    \caption{{\bf Transfer rates in the weak vibronic coupling regime associated with VAET.} ({\bf A}) Transfer rate spectra for the degenerate case ($\omega_1\!=\!\omega_2\!\equiv\!\omega$) with $(V,g_{1},g_{2})=(0.125, 0.005, 0.005)\omega$ and $\bar{n}_{0,i}\!=\!\bar{n}_{i}\!=\!0.01$. Red and green data correspond to $(\gamma_1,\gamma_2)=(0.020, 0.005)\omega$ and $(0.025, 0.075)\omega$, respectively. Blue data are the numerical results of the single-mode VAET case, where $\omega_2\!=\!g_2\!=\!\gamma_2\!=\!0$ and $\gamma_1\equiv\!\gamma\!=\!0.025\omega$.
    ({\bf B}) Transfer rate spectra for the non-degenerate case ($\omega_1>\omega_2$) with $(V,g_{1},g_{2},\omega_{2})=(0.125, 0.005, 0.005, 0.6)\omega_{1}$. Similarly, red and green data correspond to $(\gamma_1,\gamma_2)=(0.025, 0.005)\omega_1$ and $(0.025, 0.075)\omega_1$, respectively. Blue data are the same as in (A). The solid curves show the transfer rates obtained from the exponential fits of the master equation dynamics, while the point markers are their corresponding first-order FGR predictions from Eq.~\eqref{eq_kFermiVAET1}. 
    }
    \label{fig_weakcouplingVAET_S3}
\end{figure}

In contrast to the CT regime, where the collapse operators---diagonal in the Fock basis---act on displaced Fock states, the dissipation in VAET applies directly to the vibrational degrees of freedom of the eigenstates (non-displaced Fock states), and thus the associated line broadening does not require any correction factor ($C=1$). More importantly, since the first-order processes in the VAET regime only require an excitation from one of the two available vibrational modes in the two-mode system, the spectral width of the transfer resonance is exactly equal to the dissipation rate of the relevant mode, analogous to a two-level system undergoing dissipation \cite{petruccione2007}. Similarly, the number of participating phonons and their corresponding dissipation rates determine the spectral widths of the multi-phonon exchange resonances through the convolution of the Lorentzian distributions associated with the relevant vibrational levels. This is different from the broadening in CT, where the effective linewidth is independent of the amount of involved vibrational quanta and is common for all the resonances that transfer excitation from the donor to the acceptor two-dimensional potential energy wells.

Moreover, Eq.~\eqref{eq_kFermiVAET1} suggests that there is an asymmetry of the transfer rates between the resonances associated with exothermic ($E_{IF}>0$) and endothermic ($E_{IF}<0$) transfers at low temperatures, also observed in Ref.~\cite{gorman2018VAET}. However, in the very low-temperature regime ($n_{jI}\approx0$), we can further simplify the expression by considering only the terms associated with the phonon-gain resonances (exothermic transfers).  
Thus, the formula for the single-phonon transfer rates is reduced to:
\begin{eqnarray}
    k_{T}^{(1)}&=&2\pi\left|\frac{V}{2\epsilon}\right|^2 \sum_{\substack{n_{1I},n_{1F}\\n_{2I},n_{2F}}} p_{n_{1I}}p_{n_{2I}}\sum_{j=1}^2|g_j|^2\delta_{n_{kF},n_{kI}}\nonumber\\&\quad&\;\;\;\times\;(n_{jI}+1)\delta_{n_{jF},n_{jI}+1}{\rm L}(E_{IF},\gamma_j).
    \label{eq_kFermiVAET1simp}
\end{eqnarray}

In Fig.~\ref{fig_weakcouplingVAET_S3}, we plot the transfer rates given by the time constants of the master equation dynamics (solid curves) and the first-order FGR calculations (point markers). The blue data corresponds to the single-mode system, whereas the red and green data include the additional vibrational mode with weak and strong dissipation, respectively. Unlike CT, where the introduction of the second mode directly modifies the wavefunction overlaps and, in turn, the effective coupling strength between the donor and acceptor sites, the contributions of the second mode to the transfer rates in the VAET regime are different. Since the FGR predictions in Eqs.~\eqref{eq_kFermiVAET1} and ~\eqref{eq_kFermiVAET1simp} are probabilistically additive for the allowed transitions, the presence of the additional mode can increase the overall transfer rates of the system when the two vibrational modes are degenerate, as shown in Fig.~\ref{fig_weakcouplingVAET_S3}A. However, when the two modes have sufficiently different frequencies, a new transfer resonance enabled by the second mode emerges but has a negligible effect on the transfer rate associated with the first-mode resonance, which is equivalent to that of the single-mode resonance (see Fig.~\ref{fig_weakcouplingVAET_S3}B). Similar to CT, in the VAET regime, the dissipation rate of the second vibrational mode controls the trade-off between the transfer rate enhancement and increased robustness.

Although the total transfer rates are given by $k_{T} = 2\pi\left|\sum_{m=1}^\infty C_{T}^{(m)}\right|^2\rho_F(E_I)$, the complexity of the analytical expression of the transfer rates grows with the order of perturbation $m$. Since the effects of higher-order contributions are minimal due to the suppressive nature of perturbative calculations, it is therefore sufficient to consider only the numerical results of the first-order contributions for our discussion above ($k_{T} \approx k_{T}^{(1)}$). However, it is worth noting that higher-order perturbations can give rise to energy-exchange processes involving more than one phonon excitation. For instance, the second-order contributions allow two-phonon exchange processes, whose transfer rates for $|2\epsilon|\!>\!0$ are described by:
\begin{widetext}
\begin{eqnarray}
    k_{T}^{\rm (2)}\!&=&\!2\pi\left|C_T^{(2)}\right|^2\rho_F(E_I) \nonumber \\
    &=&\!2\pi\!\sum_{\substack{n_{1I},n_{1F}\\n_{2I},n_{2F}}}\!p_{n_{1I}}p_{n_{2I}}\left|\frac{V\Delta E}{\epsilon}\right|^2
    \left\{\sum_{j=1}^2\left|\frac{g_j}{2}\right|^4\delta_{n_{kF},n_{kI}} \right.\!\left[\frac{n_{jI}(n_{jI}\!-\!1)}{(2\epsilon\!+\!\omega_j)^2\omega_j^2}\delta_{n_{jF},n_{jI}-2}\!+\!
    \frac{(n_{jI}\!+\!1)(n_{jI}\!+\!2)}{(2\epsilon\!-\!\omega_j)^2\omega_j^2}\delta_{n_{jF},n_{jI}+2}\right]\!{\rm L}(E_{IF},\!2\gamma_j) \nonumber \\
    &\quad&\;+\;\left|\frac{g_1 g_2}{4}\right|^2\Biggr[(n_{1I}+1)(n_{2I}+1)\left(\frac{(2\epsilon-\omega_1)\omega_1 + (2\epsilon-\omega_2)\omega_2}{(2\epsilon-\omega_1)\omega_1(2\epsilon-\omega_2)\omega_2}\right)^2\delta_{n_{1F},n_{1I}+1}\delta_{n_{2F},n_{2I}+1} \nonumber \\
    &\quad&\;+\;(n_{1I}+1)n_{2I}\left(\frac{(2\epsilon-\omega_1)\omega_1 - (2\epsilon+\omega_2)\omega_2}{(2\epsilon-\omega_1)\omega_1(2\epsilon+\omega_2)\omega_2}\right)^2\delta_{n_{1F},n_{1I}+1}\delta_{n_{2F},n_{2I}-1} \nonumber \\
    &\quad&\;+\;n_{1I}(n_{2I}+1)\left(\frac{(2\epsilon-\omega_2)\omega_2 - (2\epsilon+\omega_1)\omega_1}{(2\epsilon+\omega_1)\omega_1(2\epsilon-\omega_2)\omega_2}\right)^2\delta_{n_{1F},n_{1I}-1}\delta_{n_{2F},n_{2I}+1} \nonumber \\
    &\quad&\;+\;n_{1I}n_{2I}\left.\left(\frac{(2\epsilon+\omega_1)\omega_1 + (2\epsilon+\omega_2)\omega_2}{(2\epsilon+\omega_1)\omega_1(2\epsilon+\omega_2)\omega_2}\right)^2\delta_{n_{1F},n_{1I}-1}\delta_{n_{2F},n_{2I}-1}\Biggl]{\rm L}(E_{IF},\gamma_1+\gamma_2)\right\}. \label{eq_kFermiVAET2}
\end{eqnarray}
\end{widetext}
For completeness, in the expression above, we include the negative energy input resonances ($2\epsilon<0$) even if they are suppressed in the case of low temperatures investigated in this work. There are two classes of two-phonon exchange processes here: single-mode exchange and double-mode exchange \cite{li2021multimodeVAET}. In the former type, two phonons from a specific vibrational mode participate in the exchange, while the latter class consists of processes in which one phonon from each mode contributes to the two-phonon-assisted processes.

Another difference between the CT and VAET regimes is related to high-$\Delta E$ transfers. In the CT regime, both the initial temperature and displacement-dependent Franck-Condon factors determine whether highly excited processes are allowed. However, in the VAET regime, multi-phonon exchange processes are always suppressed due to the progressively decreasing coupling strengths with respect to the orders of perturbative calculations.

Alternatively to the perturbation analysis, we can use the non-interacting blip approximation to track the spin dynamics of VAET, as described below in Appendix \ref{SM_niba}. Despite its closed-form formula, its evaluation heavily relies on the complexity of the spectral density function describing the environmental influences on the pure spin systems.

\section{Solution for VAET under NIBA}
\label{SM_niba}

In this section, we derive a closed-form solution for open-system VAET under the non-interacting blip approximation (NIBA) by considering the following spin-boson model \cite{leggett1987dissipative,dekker1987blip}:
\begin{equation}
    H_{\rm sb}=\frac{\Delta E}{2}\sigma_z+V\sigma_x+\sum_{k=1}^{\infty}{\frac{\lambda_k}{2}\sigma_z\left(a_k^\dagger+a_k\right)}+\sum_{k=1}^{\infty}{\omega_ka_k^\dagger a_k},
    \label{eq_sp_mod}
\end{equation}
where, in the continuum limit, the couplings between the spin and each bosonic mode and the frequencies of the infinite bosonic bath are characterized by the spectral density $J\left(\omega\right)=\pi\sum_{k}{\lambda_k^2\delta(\omega-\omega_k)}$. It has been shown in Ref.~\cite{lemmer2018engineering} that if the spectral density takes the form:
\begin{equation}
    J(\omega) = \sum_{i=1}^2g^2_i\left(\frac{\gamma_i}{\gamma_i^2+\left(\omega_i-\omega_{im}\right)^2}-\frac{\gamma_i}{\gamma_i^2+\left(\omega_i+\omega_{im}\right)^2}\right),
    \label{eq_lor_den}
\end{equation}
with $\gamma_i,\;\omega_i,$ and $g_i$ defined in the main text and $\omega_{im}=\sqrt{\omega_i^2-\gamma_i^2}$, the spin dynamics generated by the spin-boson model in Eq.~\eqref{eq_sp_mod} is equivalent to that generated by the dissipative LVCM system described by Eq.~\eqref{eq_master} in the main text under the conditions $\gamma_i\ll\omega_{im}$ and $\gamma_i\ll k_BT_i$ for $k_BT_i\equiv k_BT=1/\beta$ (the same temperature for all bosonic bath modes).

Given the above spectral density function, the coupling strength between the spin and each bosonic bath mode is characterized by $g_i$, which is small when compared to $\omega_i$, $\Delta E$, and $V$ in the VAET regime. Therefore, it is appropriate to apply the weak-coupling approximation with NIBA \cite{leggett1987dissipative,dekker1987blip}. The vibronic coupling terms in Eq.~\eqref{eq_sp_mod} can then be canceled out by applying a unitary transformation $U=\exp{(\sum_{k}{\alpha_k(a_k^\dagger-a_k)})}$ with $\alpha_k=-\frac{1}{2}\frac{\lambda_k}{\omega_k}\sigma_z$, such that the transformed Hamiltonian becomes:
\begin{equation}
    U^\dagger HU=\frac{\Delta E}{2}\sigma_z+V\left(\sigma^+e^{-iB}+\sigma^-\sigma^{iB}\right)+\sum_{k}{\omega_ka_k^\dagger a_k},
\end{equation}
where the operator $B=i\sigma_z\sum_{k}{\frac{\lambda_k}{\omega_k}(a_k^\dagger-a_k)}$. We then follow the procedure in the Appendix of Ref.~\cite{kamar2024dephasing} and write down the spin equations of motion in the Heisenberg picture:
\begin{eqnarray}
    \partial_t\sigma_{z,H}&=&i2V(\sigma_H^-e^{iB(t)}-\sigma_H^+e^{-iB\left(t\right)}),\nonumber\\
    \partial_t\sigma_H^+&=&i\Delta E\sigma_H^+-iV\sigma_{z,H}e^{iB\left(t\right)},
\end{eqnarray}
where $\sigma_{z,H}$ and $\sigma_H^+$ are the time-dependent Pauli operators in the Heisenberg picture. We can solve for $\sigma_H^+$ in terms of $\sigma_{z,H}$ from the second equation and substitute the solution into the first equation. By taking the expectation value, we get:
\begin{eqnarray}
    \partial_t\langle\sigma_z\rangle&=&-2V^2\int_{0}^{t}{ds \langle\sigma_z\rangle\left(s\right) e^{-iE\left(t-s\right)} \langle e^{-iB\left(s\right)}e^{iB\left(t\right)}\rangle}\nonumber \\ &\quad&\quad \quad\quad\quad+ \; {\rm h.c.}
    \label{eq_sigma_z}
\end{eqnarray}
With NIBA, we suppose that the bath evolution is decoupled from that of the spin such that the expectation value $ \langle e^{-iB\left(s\right)}e^{iB\left(t\right)}\rangle$ can approximately be calculated by considering the evolution of the bosonic modes under the free bath Hamiltonian $\sum_{k}{\omega_ka_k^\dagger a_k}$. Assuming each bath mode is at equilibrium with a temperature $k_BT$ and utilizing the second-order cumulant expansion $\langle\exp{X}\rangle\rightarrow\exp{\langle X+\frac{1}{2}\text{Var}\left(X\right)\rangle}$, we have:
\begin{eqnarray}
\langle e^{-iB\left(s\right)}e^{iB\left(t\right)}\rangle&=&\exp{\left(-iQ_1\left(t-s\right)\right)}\exp{\left(Q_2\left(t-s\right)\right)},\nonumber\\
\label{eq_B_exp}
\end{eqnarray}
where 
\begin{eqnarray}
    Q_1\left(\tau\right)&=&\frac{1}{\pi}\int_{-\infty}^{\infty}{d\omega J(\omega)\sin{\left(\omega\tau\right)}/\omega^2}, \nonumber\\
    Q_2\left(\tau\right)&=&\frac{1}{\pi}\int_{-\infty}^{\infty}{d\omega J(\omega)\left(\cos{\left(\omega\tau\right)}-1\right)\coth{\left(\beta\omega/2\right)}/\omega^2}.\nonumber\\
\end{eqnarray}
By inserting Eq.~\eqref{eq_B_exp} into Eq.~\eqref{eq_sigma_z}, the equation of motion for $\langle\sigma_z\rangle$ takes the form of a convolution with a kernel function $f(\tau)$:
\begin{eqnarray}
\partial_t\left\langle\sigma_z\right\rangle&=&\int_{0}^{t}dsf\left(t-s\right)\left\langle\sigma_z\right\rangle\left(s\right),\nonumber\\
f\left(\tau\right)&=&-4V^2\cos{\left(Q_1\left(\tau\right)-\Delta E\tau\right)}e^{Q_2(\tau)},
    \label{eq_conv}
\end{eqnarray}
which can be formally solved using the Laplace transform:
\begin{equation}
    \langle\sigma_z\rangle\left(t\right)=\mathcal{L}_T^{-1}\left[\frac{\langle\sigma_z\rangle(t=0)}{\zeta-\widetilde{f}(\zeta)}\right],
    \label{eq_lap_sol}
\end{equation}
where $\tilde{f}(\zeta)=\mathcal{L}_T[f(\tau)]$ is the Laplace transform of the kernel function $f$.

\section{Interference effects in two-mode VAET}
\label{SM_coherence}

To better understand the interference effects in two-mode VAET systems leading to the enhancement at $\omega_1 + \omega_2$ resonance in the main text, we examine the zero-temperature case ($p_{n_{jI}=0}=1,\; p_{n_{jI}\neq0}=0$) in Eq.~\eqref{eq_kFermiVAET2}, which gives the rates of transfer between the eigenstates in Eq.~\ref{eq_VAETpertsystem} for $\Delta E = \sqrt{(\omega_1+\omega_2)^2-V^2} \equiv E_{\rm dual}$ and $\Delta E = \sqrt{(2\omega_i)^2-V^2} \equiv E_{\rm single}^i$: 
\begin{eqnarray}
    k_T^{(2)}(E_{\rm dual})&=&2\pi\left[\frac{g_1g_2VE_{\rm dual}}{(\omega_1+\omega_2)^2}\left(\frac{1}{\omega_1}+\frac{1}{\omega_2}\right)\right]^2
    \nonumber\\
    &\quad&\quad\quad\quad\times\;{\rm L}(E_{\rm dual},\gamma_1+\gamma_2),\label{eq_Edual}\\ 
    k_T^{(2)}(E_{\rm single}^i)&=&2\pi\left[\frac{g_i^2VE_{\rm single}^i}{(2\omega_i)^2}\frac{\sqrt{2}}{\omega_i}\right]^2
    \nonumber\\
    &\quad&\quad\quad\quad\times\;{\rm L}(E_{\rm single}^i,2\gamma_i), \label{eq_Esingle}
\end{eqnarray}
respectively. When $\omega_1=\omega_2\equiv\omega$, $g_1=g_2\equiv g$, and $\gamma_1=\gamma_2\equiv \gamma$, Eqs.~\eqref{eq_Edual} and \eqref{eq_Esingle} give $k_T^{(2)}(E_{\rm dual})=2\times k_T^{(2)}(E_{\rm single}^i)$, explaining the enhancement at the second resonance of the degenerate two-mode VAET compared to its single-mode counterpart, as observed in Fig.~\ref{fig_VAETboth}A. It also suggests that the dual-phonon process associated with the $\omega_1+\omega_2$ energy input accounts for approximately half of this enhancement. Similarly, for $\omega_1>\omega_2$ and $g_i\ll\omega_i$, we get $k_T^{(2)}(E_{\rm dual})\!>\!k_T^{(2)}(E_{\rm single}^i)$, as observed in Fig.~\ref{fig_VAETboth}C.

%----------------------------
\begin{figure}[t!]
    \centering
    \includegraphics[width=1\linewidth]{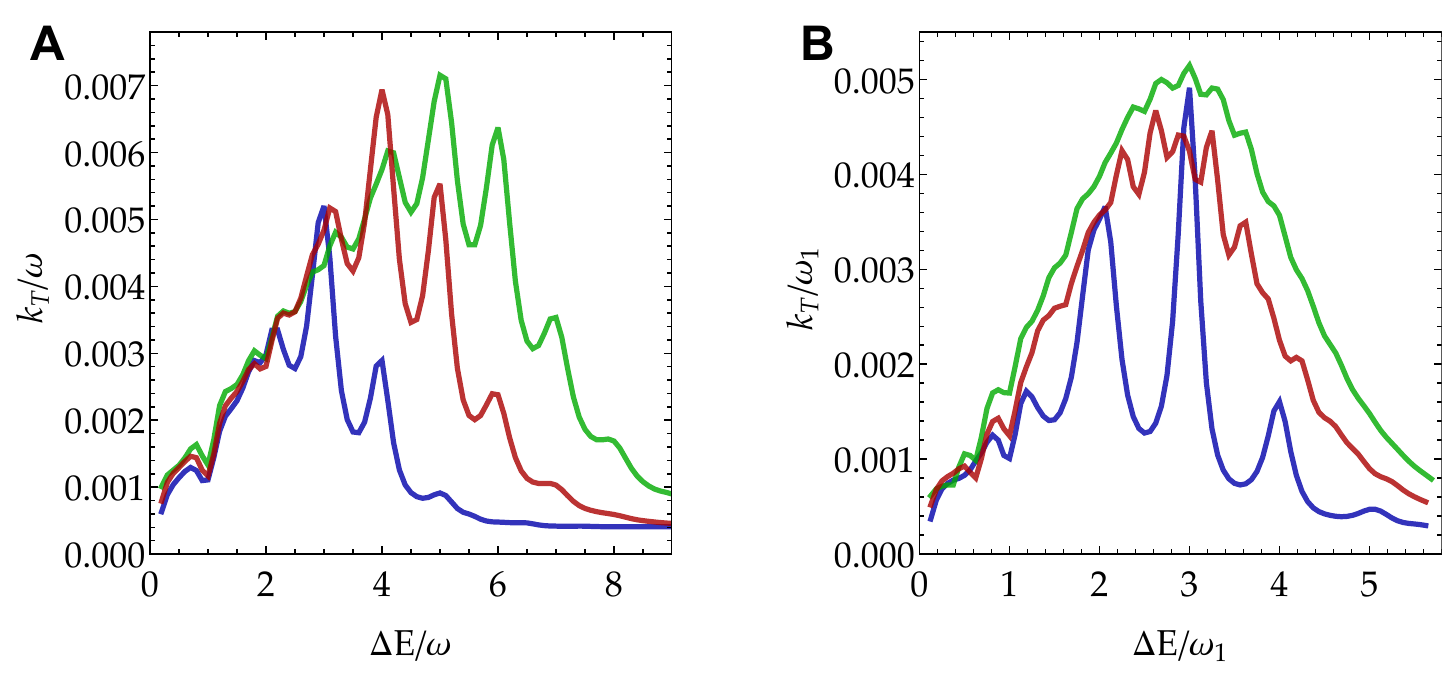}
    \caption{{\bf Transfer rates for different numbers of vibrational modes involved in CT.} ({\bf A}) Transfer rate spectra for the degenerate case ($\omega_1=\omega_2=\omega_3\equiv\omega$) with $(V,g_{i},\gamma_{i})=(0.2, 1, 0.02)\omega$ for $i=1,2,$ and $3$.
    ({\bf B}) Transfer rate spectra for the non-degenerate case ($\omega_1\!>\!\omega_2\!>\!\omega_3$) with $(V,g_{1},\gamma_{1},\omega_{2},\omega_3)=(0.125, 1, 0.025, 0.625, 0.375)\omega_{1}$, $(g_{2},\gamma_{2})=(1,0.020)\omega_2$, and $(g_{3},\gamma_{3})=(1,0.033)\omega_3$.  In both cases, the solid curves represent the transfer rates calculated from Eq.~\eqref{eq_master} using the definition in Eq.~\eqref{eq_lifetime_kT} with the same additional decoherences used in the main text. The blue, red, and green curves are the numerical results of the CT systems with one vibrational mode ($i=1$ only), two vibrational modes ($i=1$ and $2$), and three vibrational modes ($i=1,2,$ and $3$), respectively.
    }
    \label{fig_manymodesET}
\end{figure}
%----------------------------

\section{Beyond two-mode LVCM} \label{SM_threemode}

\begin{figure}[t!]
    \centering
    \includegraphics[width=1\linewidth]{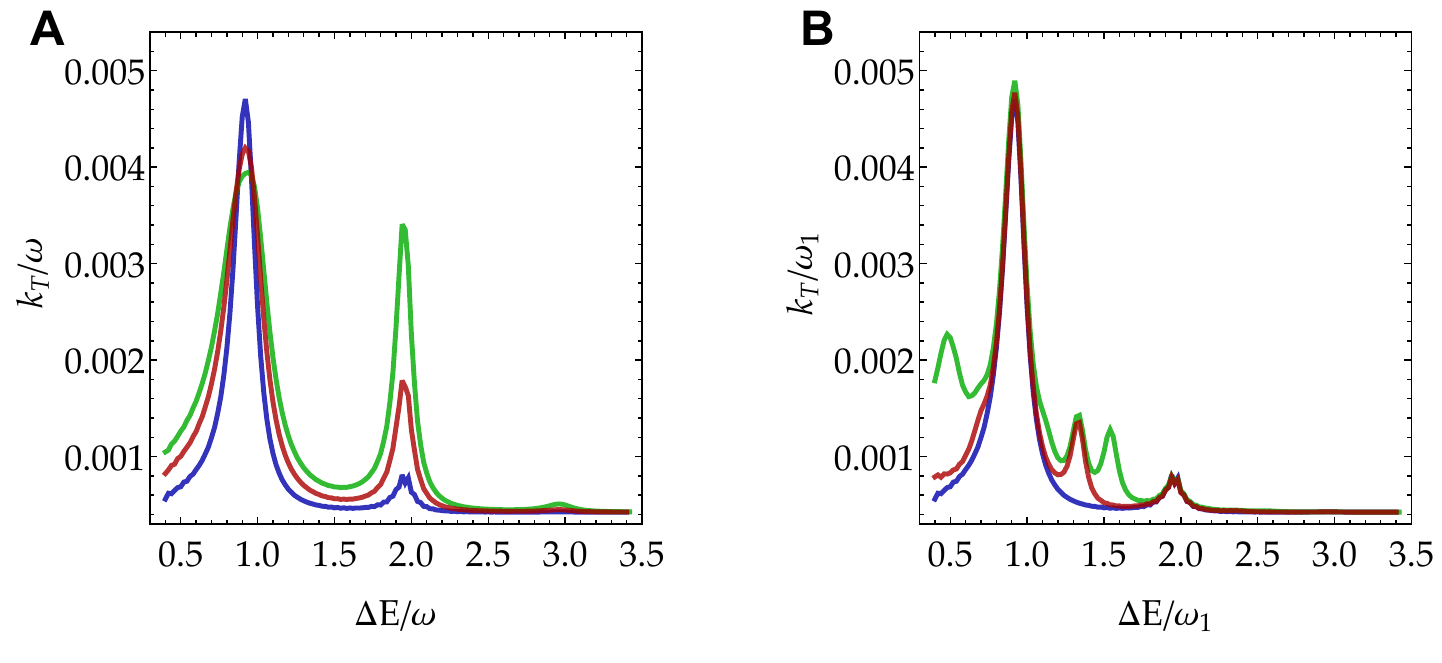}
    \caption{{\bf Transfer rates for different numbers of vibrational modes involved in VAET.} ({\bf A}) Transfer rate spectra for the degenerate case ($\omega_1=\omega_2=\omega_3\equiv\omega$) with $(V,g_{i},\gamma_{i})=(0.2, 0.2, 0.04)\omega$ for $i=1,2,$ and $3$.
    ({\bf B}) Transfer rate spectra for the non-degenerate case ($\omega_1>\omega_3>\omega_2$) with $(V,g_{1},\gamma_{1},\omega_{2},\omega_3)=(0.2, 0.2, 0.04, 0.4, 0.6)\omega_{1}$, $(g_{2},\gamma_{2})=(0.2,0.04)\omega_2$, and $(g_{3},\gamma_{3})=(0.2,0.04)\omega_3$. In both cases, the solid curves represent the transfer rates calculated from Eq.~\eqref{eq_master} using the definition in Eq.~\eqref{eq_lifetime_kT} with the same additional decoherences used in the main text. The blue, red, and green curves are the numerical results of the VAET systems with one vibrational mode ($i=1$ only), two vibrational modes ($i=1$ and $2$), and three vibrational modes ($i=1,2,$ and $3$), respectively.
    }
    \label{fig_VAETmanymodes}
\end{figure}

Here, we show how the conclusions of our work on the two-mode models can be straightforwardly extended to the three-mode models and presumably to multi-mode models with $i>3$ with Figs.~\ref{fig_manymodesET} and \ref{fig_VAETmanymodes}. Similar to the degenerate two-mode CT process, the presence of the additional vibrational mode in the degenerate three-mode CT case increases the number of state configurations on the hybridized energy surfaces (now three-dimensional), which leads to a wider $\Delta E$ region with monotonically increasing transfer rates and sharper peaks at high-$\Delta E$ ($\ge 5\omega$) resonances corresponding to the release of the initially trapped population in the three-dimensional upper adiabatic states (see Fig.~\ref{fig_manymodesET}A). Meanwhile, as shown in Fig.~\ref{fig_manymodesET}B, when the vibrational energies are not all equal, the transfer rate spectrum features a smooth transfer profile similar to that of two-mode CT except for the slightly increased rates, which can be explained by the additional transfer channels provided by the third vibrational mode.

In the case of three-mode VAET, the vibrational degrees of freedom provide access to three oscillator baths from which units of different phonon energies can be taken to assist the transfer and thus enable many combinative pathways for transfer resonances to occur, as shown in Fig.~\ref{fig_VAETmanymodes}B. With degeneracy across the vibrational modes, the linear combinations of energy supply for the two-phonon resonance ($2\omega_1 = 2\omega_2 = 2\omega_3 = \omega_1 + \omega_2 = \omega_1 + \omega_3 = \omega_2 + \omega_3 = 2\omega$) drastically increase the transfer rates beyond the two-mode VAET ($2\omega_1 = 2\omega_2 = \omega_1 + \omega_2 = 2\omega$). However, compared to two-mode VAET, there is again a slight decrease in the transfer rate of the first resonance due to the extra broadening caused by the dissipation of the additional mode (see Fig.~\ref{fig_VAETmanymodes}A).

While the same intuition on the two-phonon resonances can be applied to the third and higher-order resonances for expected increases in the transfer rates, low-temperature VAET systems limit the transfers to few-phonon-assisted processes. As concluded in the main text, the numerical results for the three-mode systems support the generalization of the roles of mode degeneracy and vibronic coupling strength in two-mode LVCM to two-site systems with a higher number of vibrational modes than 2 in both phenomenological regimes (CT and VAET).

\section{VAET-CT crossover}
\label{SM_crossover}

%----------------------------
\begin{figure*}[!]
    \centering
    \includegraphics[width=0.9\linewidth]{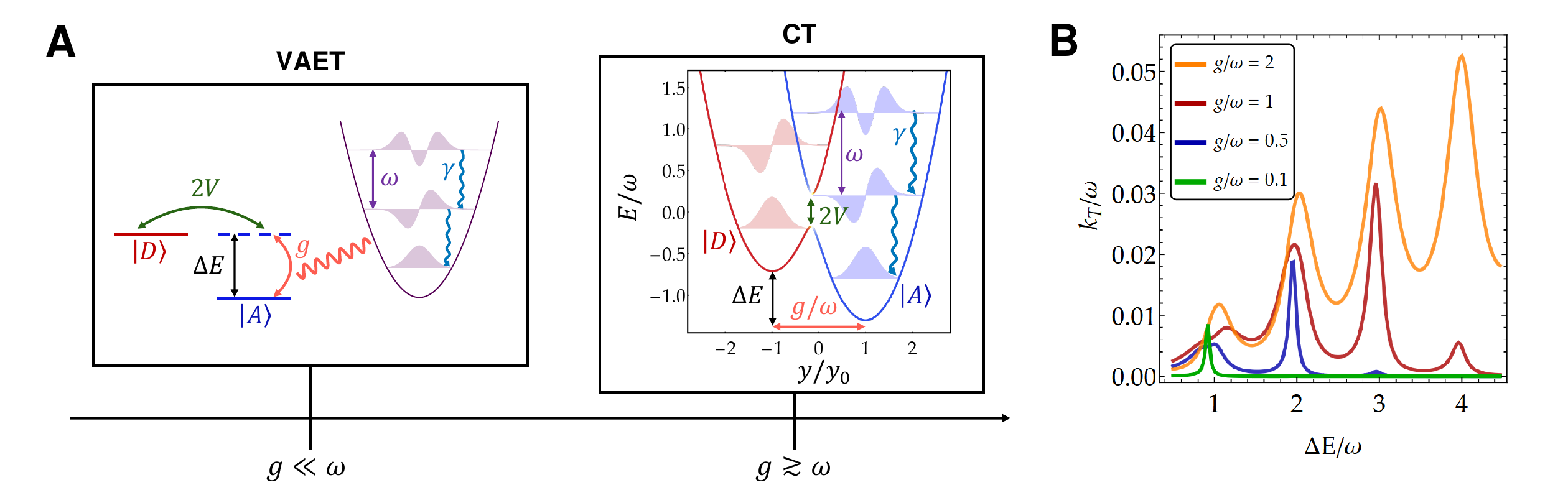}
    \caption{{\bf VAET-CT crossover in single-mode LVCM.} ({\bf A}) Schematic diagram illustrating the two types of transfer dynamics in single-mode models. ({\bf B}) Transfer rate spectra with $(V,\gamma_{i})=(0.2, 0.05)\omega$ at various values of $g/\omega$.
    All curves show the transfer rates calculated from Eq.~\eqref{eq_master} using the definition in Eq.~\eqref{eq_lifetime_kT}.
    }
    \label{fig_crossover}
\end{figure*}
%----------------------------

In this section, we investigate a low-temperature single-mode LVCM system with all the system parameters fixed except for the vibronic coupling strength to understand the role of displacement in transitioning the transfer dynamics between distinct characteristic regimes, such as VAET and CT (see Fig.~\ref{fig_crossover}A). For consistency, we choose the electronic coupling strength to be larger than the dissipation rate. We compare the transfer rate spectra for different values of $g/\omega=\{0.1, 0.5, 1, 2\}$ in Fig.~\ref{fig_crossover}B.

\begin{figure}
    \centering\includegraphics[width=1\linewidth]{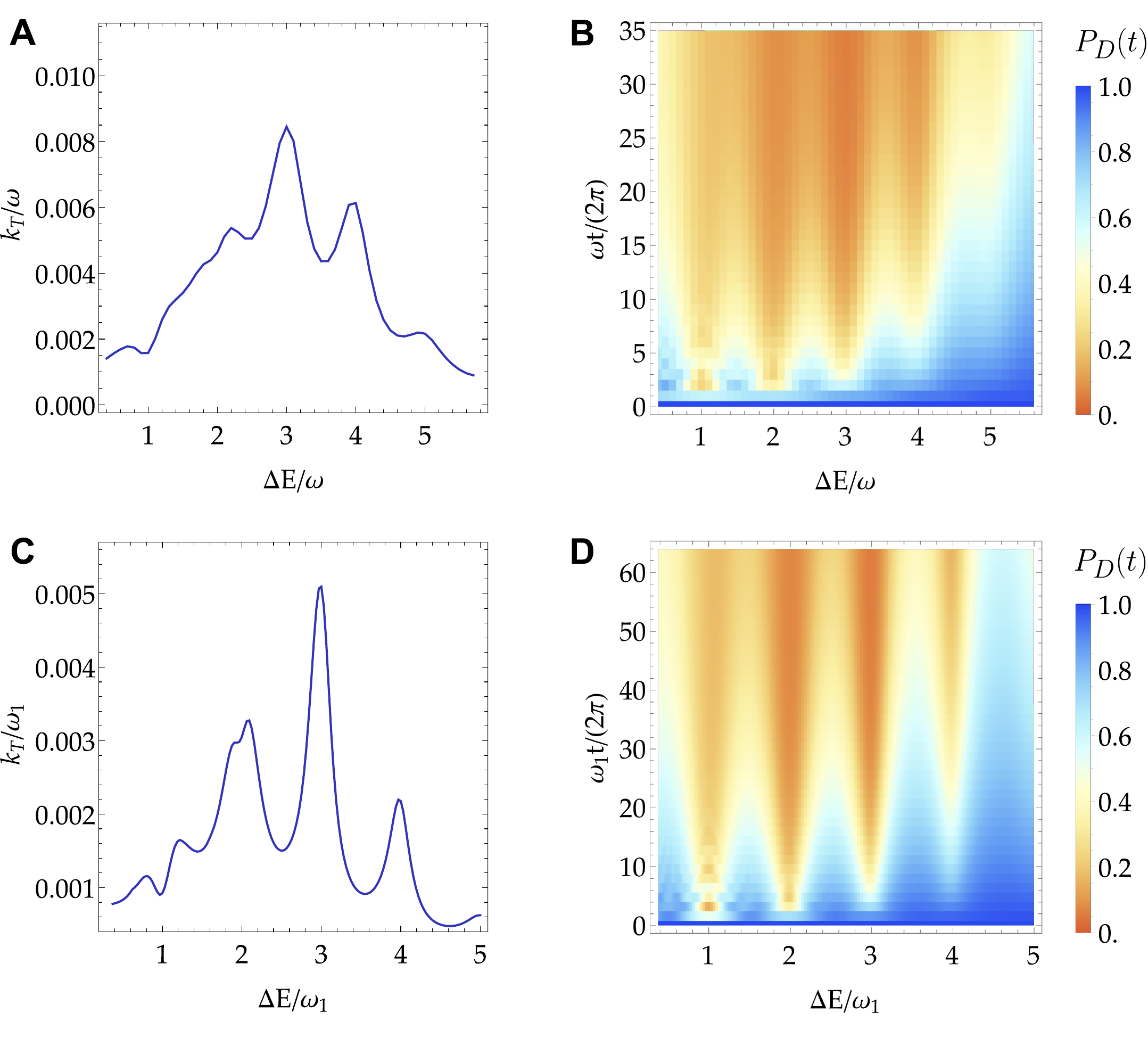}
    \caption{{\bf CT rates for single-mode systems.} ({\bf A}) Transfer rate spectrum with $(V,g_{1},\gamma_{1})=(0.200, 1.200, 0.036)\omega$. The solid blue curve shows the transfer rate calculated from Eq.~\eqref{eq_master}, using the definition in Eq.~\eqref{eq_lifetime_kT} and including spin decoherence ($\gamma_{z}\!=\!0.0014 \omega$) and motional dephasing ($\gamma_{1m}\!=\!0.0160\omega$).
    ({\bf B}) Numerical donor population evolution $P_D(t)$ versus energy gap $\Delta E$ and the number of vibrational oscillations $\omega t/2\pi$, using the same parameters as the solid red curve in {(A)}. ({\bf C and D}) Same layout as in {(A and B)} with $(V,g_{1},\gamma_{1},\gamma_{z},\gamma_{1m})=(0.138, 1.029, 0.023, 0.0009, 0.0100)\omega_{1}$.}
    
    \label{fig_CT1M}
\end{figure}

At small $g/\omega=0.1$, the system undergoes VAET dynamics, showing the expected sharp transfer resonance at $\Delta E = \sqrt{\omega^2-(2V)^2}$ from the single-phonon assisted process. As $g/\omega$ increases, the vibrational state of the system becomes effectively displaced, giving rise to non-negligible transfer rates at resonances associated with higher-order phonon-assisted processes. However, when $g\sim\omega$, the displaced LVCM system becomes a CT system, particularly an adiabatic CT system with $|V|\sim \lambda/4$, leading to monotonically increasing transfer rates at low $\Delta E$ and sharp resonances at higher $\Delta E$ (see the results with $g/\omega = 0.5$ and 1 in Fig.~\ref{fig_crossover}B). From these observations, we infer that $g\approx 0.5\omega$ marks the crossover between VAET and CT, where the first resonant peak associated with a single-phonon assisted process is effectively washed out into a monotonically increasing feature in the transfer rate. Therefore, in this work, we associate systems with $g_i/\omega_i\lesssim0.5$ to non-perturbative VAET and systems with $g_i/\omega_i\gtrsim1$ to non-perturbative CT, all with strong electronic coupling ($|V|\sim\lambda_i/4$).

As we further increase $g/\omega$, the system progressively evolves into the nonadiabatic regime of CT with $|V|<\lambda/4$, where discernible resonance peaks are recovered across the transfer rate spectrum. Different from VAET, these resonances occur at integer multiples of $\omega$, and the position of the strongest resonance depends on the displacement between the donor and acceptor potential energy surfaces. The two distinct regimes within CT, nonadiabatic and adiabatic transfers, are determined by the relative strengths of the electronic coupling to the reorganization energy of the system and the dissipation rate of the vibrational mode, which have been studied in Refs.~\cite{schlawin2021electrontransfer,so2024electrontransfer}.

\section{Single-mode charge transfer dynamics}
\label{SM_CT1M}

Fig.~\ref{fig_CT1M} shows the transfer rate spectra and the corresponding donor population dynamics from numerical simulations for the single-mode cases, which are used for comparison with the two-mode CT results in Fig.~\ref{fig_CT} of the main text.

%----------------------------
\end{document}